\documentclass[useAMS,usenatbib,usegraphicx]{mn2e}
\usepackage{amssymb,amsmath,amsfonts,graphicx}
\usepackage{epsfig,color} 

\def\Msun{\hbox{$\rm\, M_{\odot}$}}

\title[Satellite radial distribution]{Satellite abundances around bright
  isolated galaxies II: radial distribution and environmental effects}

\author[Wang et al.]{Wenting Wang$^{1,2,3}$, Laura V. Sales$^{1,4}$ , Bruno M. B. Henriques$^{1}$ and Simon White$^{1}$\\
  {}$^{1}$Max Planck Institut fur Astrophysik, Karl-Schwarzschild-Str. 1, 85741 Garching b. M\"unchen, Germany\\
  {}$^{2}$Key Laboratory for Research in Galaxies and Cosmology of Chinese Academy of Sciences, Max-Panck-Institute Partner\\
    Group, Shanghai Astronomical Observatory, Nandan Road 80, Shanghai 200030, China\\
  {}$^{3}$Institute for Computational Cosmology, University of Durham, South Road, Durham, DH1 3LE, UK\\
  {}$^{4}$Harvard-Smithsonian Center for Astrophysics, 60 Garden Street, Cambridge, MA, 02138, USA
}
\begin{document}



\maketitle

\begin{abstract}
  We use the SDSS/DR8 galaxy sample to study the radial distribution
  of satellite galaxies around isolated primaries, comparing to
  semi-analytic models of galaxy formation based on the Millennium and
  Millennium-II simulations. SDSS satellites behave differently around
  high- and low-mass primaries: those orbiting objects with $M_*
  >10^{11}M_\odot$ are mostly red and are less concentrated towards
  their host than the inferred dark matter halo, an effect that is
  very pronounced for the few blue satellites.  On the other hand,
  less massive primaries have steeper satellite profiles that agree
  quite well with the expected dark matter distribution and are
  dominated by blue satellites, even in the inner regions where strong
  environmental effects are expected. In fact, such effects appear to
  be strong only for primaries with $M_* > 10^{11}M_\odot$.  This
  behaviour is not reproduced by current semi-analytic simulations,
  where satellite profiles always parallel those of the dark matter
  and satellite populations are predominantly red for primaries of all
  masses.  The disagreement with SDSS suggests that environmental
  effects are too efficient in the models.  Modifying the treatment of
  environmental and star formation processes can substantially
  increase the fraction of blue satellites, but their radial
  distribution remains significantly shallower than observed. It seems
  that most satellites of low-mass primaries can continue to form
  stars even after orbiting within their joint halo for 5 Gyr or more.

\end{abstract}


\section{Introduction}
\label{sec:intro}

Satellite galaxies can contribute substantially to our understanding
of galaxy formation. In the current structure formation paradigm,
galaxies form by the cooling and condensation of gas at the centres of
an evolving population of dark matter halos that are an order of
magnitude larger in both mass and linear size than the visible
galaxies \citep{White_Rees1978}.  Comparable contributions to the
growth of such halos come from smooth accretion of diffuse matter and
from mergers with other halos spread over a very wide range in mass
\citep{Wang2011}. The more massive accreting halos will normally have
their own central galaxies, and after infall these become
``satellites'' of the galaxy at the centre of the dominant halo,
orbiting it within their own ``subhalos''.  Later, the satellites may
merge with the central galaxy and so contribute to its growth.

High-resolution cosmological simulations predict not only the masses,
positions and velocities of dark matter halos but also those of the
subhalos they contain \citep[e.g.][]{Moore1999, Gao2004a, Springel2008,
Gao2008,Gao2011}. Linking such data over time then allows 
construction of the assembly history of every system in the simulated 
volume. In combination with a model for galaxy formation, such 
halo/subhalo merger trees can be used to predict the development of 
the full galaxy population in the region considered. This can be 
compared directly with properties of observed populations such as 
abundances, scaling relations, clustering and evolution 
\citep[e.g.][]{Springel2001,Bower2006,Croton2006,Guo2011}. A particular 
strength of such ``semi-analytic'' population simulations is that 
they enable evaluation of the relative sensitivity of these observables 
to cosmological and to galaxy formation parameters \citep[e.g.][]{Wang2008,
Guo2013}. Satellite galaxies play an important role in such work
because they are particularly sensitive to environmental effects and
to the assembly history of halos.

In \citet[][ PaperI hereafter]{Wang_White2012} we used the Sloan
Digital Sky Survey (SDSS) to study the luminosity, mass and colour
distributions of satellite galaxies as a function of the properties of
their host. A comparison of our observational results to semi-analytic
galaxy formation simulations within the concordance $\Lambda$CDM
cosmology showed good overall agreement for satellite abundances,
inspiring some confidence in the realism of the particular galaxy
formation model used \citep[from][]{Guo2011}, but large discrepancies
for satellite colour distributions confirmed earlier demonstrations
that such models substantially overestimate the environmental
suppression of star formation \citep[e.g.][]{Font2008,
  Weinmann2009}. In this paper we extend our earlier work through a
detailed analysis of the radial distribution of satellites around
their hosts. This enables further exploration both of the successes
and of the failures of the galaxy formation model.

The observational study of satellite number density profiles benefited
enormously from the advent of wide-angle spectroscopic surveys such as
the Two Degree Field Galaxy Redshift Survey
\citep[2dFGRS,][]{Colless2001} and the SDSS \citep{York2000}. The
availability of redshift measurements for almost all objects above
some apparent magnitude limit allows the full three-dimensional
distribution of objects to be studied (although in ``redshift space''
rather than true position space) greatly facilitating the
identification of host/satellite systems.
Several studies concluded that the mean radial satellite distribution
in such spectroscopic samples can be fit (in projection) by a
power-law $\Sigma_{\rm sat} \propto r^{-\alpha}$, although the range
of indices quoted is quite broad $\alpha \sim 0.9$ to $1.7$
\citep{Sales2005,vandenBosch2005,Chen2006, Chen2008}. There is some
indication that this index correlates with the properties of the
primaries and/or satellites under consideration, but results are also
rather noisy because of the relatively bright lower limit on the
luminosity of the satellites which is enforced by the spectroscopic apparent
magnitude limit.

In addition to being restricted to relatively bright objects,
spectroscopic satellite samples are also subject to selection effects
such as redshift incompleteness due to fibre-fibre collisions and
survey geometry constraints which particularly affect their coverage
of close pairs. In this context, photometric samples offer an
interesting alternative, since they are complete at all separations
and to apparent magnitude limits which are typically $3-4$ magnitudes
fainter than the corresponding spectroscopic surveys. For example, the
SDSS/DR8 data are effectively complete to $r$-band magnitudes
$m_r=17.7$ and $m_r=21$ for the spectroscopic and photometric
catalogues, respectively \citep{Aihara2011}.

Inspired by this, several groups have recently analyzed
primary/satellite samples, where the primary galaxies are selected
from spectroscopic surveys, ensuring their distances and environments
are well characterized, but their satellite populations are identified
in deeper photometric data and so must be corrected statistically for
the inevitable foreground and background contamination
\citep[e.g. ][]{Lares2011, GuoQuan2012, Nierenberg2011,
  Nierenberg2012, Jiang2012,Tal2012}. This approach is reminiscent of
the pioneering work in this field, where satellites were identified on
photographic plates around relatively bright primary samples
\citep{Holmberg1969, Lorrimer1994}.  The projected satellite profiles
measured in these hybrid studies are also consistent with power-laws
$\Sigma_{\rm sat} \propto r^{-0.9,-1.2}$, and again correlations are
seen between the slope of the profiles and the colour/mass/type of the
primaries and satellites.

Despite this superficial agreement, there are large discrepancies
between recently published studies of satellite radial distributions. 
Some authors find satellite profiles to be steeper than the NFW profile 
\citep{Navarro1996,Navarro1997} predicted for the dark matter 
\citep{Watson2012,Tal2012,GuoHong2014}; others consider them as good 
tracers of the dark matter \citep{Nierenberg2012}; yet others find them 
to be less concentrated than the dark halos they inhabit \citep{Budzynski2012,
  Wojtak2013}. The trends found with intrinsic properties of 
satellites/primaries also disagree between studies. For instance,
whereas Watson et al. and Tal et al. find that bright satellites have
steeper profiles, Guo et al. and Budzynski et al. conclude that faint
companions are more strongly concentrated.  Nierenberg et al. find no
variation in profile slope with satellite mass.

Some of this disagreement can plausibly be traced to differing sample
definitions. For example, \cite{Tal2012} studied satellite profiles
around Luminous Red Galaxies (LRGs) at $0.28<z<0.4$, whereas both
\cite{Watson2012} and \cite{GuoQuan2012} used SDSS Main Sample
galaxies at lower redshifts.  The redshift range probed by
\cite{Nierenberg2012} is $0.1<z<0.8$, based on data from the deeper
COSMOS survey, but their samples are relatively small so trends may be
masked by counting noise. Furthermore, \cite{Tal2012} and
\cite{Watson2012} studied satellite radial profiles down to very small
separations ($r_p\sim 30$kpc) and their inference of a steeper than
NFW distribution depends on these scales. Careful photometric
corrections are needed in such work, because satellite magnitudes are
systematically biased by their proximity to a much brighter central
galaxy. Deblending and background estimation effects can be
substantial in this situation and are quite uncertain
\citep[e.g.][]{Mandelbaum2006}. Not all authors apply such corrections
\citep[e.g.][]{GuoQuan2012} and in consequence their results at the
smallest separations may be compromised.

Variations in the radial distribution of satellites as a function of
primary or satellite properties provide important clues to the
processes driving galaxy evolution, in particular, to the influence of
environmental effects. Tidal disruption and ram-pressure stripping are
believed to be the main agents of structural change in satellites once
they have fallen into their host halos. Extended reservoirs of gas may
be removed, causing the satellites to run out of fuel for star
formation, or gas and stars may be removed directly from the visible
regions of the galaxies. As a result, satellites are predicted to be
less active and redder than otherwise similar galaxies in the field.

There is clear observational evidence for effects of this kind.
Studies of galaxy correlations show enhanced clustering of red objects
at fixed stellar mass \citep[see e.g.][]{LiCheng2006,
Zehavi2011} and there is a consensus among authors that
the fraction of red and passive satellite galaxies is larger than for
central galaxies of similar mass \citep[see e.g. ][]{vandenBosch2008,
  Yang2009, Weinmann2009}.  There are also, however, clear indications
that this increased red fraction among satellites is a function of the
stellar or halo mass of the primary, suggesting that environmental
effects are weak or even negligible for satellites orbiting low-mass
primaries \citep[e.g.][]{Weinmann2006a,Prescott2011,Wetzel2012}.

Theoretical predictions based on semi-analytical models successfully
reproduce several of these trends, but typically overproduce
the fraction of red satellites \citep[e.g.][]{Coil2008}. Recent improvements
in the modelling of gas removal and tidal stripping have improved the
situation \citep{Font2008,Guo2011}, but a significant problem still
persists \citep{Weinmann2011}.  The radial distributions of red and
blue satellites and their relation to the properties of the primary
galaxy give additional information about environmental influences on
satellites, complementing the information provided by the relative
abundances of the two populations.

Despite difficulties in matching the observed colour distribution,
simulations have proven useful for interpreting the observed
properties of satellite galaxies. \citet{Kravtsov2004} and
\citet{Gao2004b} used N-body simulations to argue that the observed
radial distribution of luminous satellites is more easily understood
if these objects populate the most massive subhalos at the time of
infall, rather than the most massive today. The distribution of
luminous satellites in both hydrodynamical and semi-analytic
simulations suggests that they may be reasonable tracers of the
underlying dark matter distribution of their host halo
\citep[e.g. ][]{Gao2004b,Nagai2005,Sales2007c}. Numerical simulations
also show that the time of infall of satellites onto their host is
correlated with their current distance from halo centre
\citep[e.g.][]{Gao2004a}, a relation that becomes tighter if we
consider satellite orbital binding energy \citep{Rocha2012}.  Thus the
radial distribution of satellites encodes information about the
assembly of dark matter halos that is not otherwise observationally
accessible.

In this paper we study the radial distribution of satellites in a
hybrid primary/satellite sample selected from the spectroscopic +
photometric SDSS/DR7 and DR8 catalogues.  We go beyond previous work
by comparing our results with a mock-galaxy catalogue generated from
the Millennium and Millennium-II simulations \citep{Springel2005a,
  Boylan-Kolchin2009} using the semi-analytical model of
\citet{Guo2011}. The mock sample allows an improved assessment of the
projection and sample selection effects, facilitating the physical
interpretation of the observed profiles. At the same time, we are able
to test the galaxy formation model by contrasting its predictions with
observables it was not tuned to reproduce. This paper follows
naturally from the analysis presented in Paper I which focused on the
abundance and mass spectrum of satellites around isolated primaries.

This paper is organized as follows: our data sources and the selection
criteria we apply to observed and simulated catalogues are described
in Sec.~\ref{sec:data}. We report the trends found in the radial
distribution of satellites according to primary/satellite colours and
masses in Sec.~\ref{ssec:pricolor} and \ref{ssec:satcolor}, while we
discuss the implications for environmental modulation of star formation
in Sec.~\ref{sec:tinf}.  We summarize and discuss our main conclusions
in Sec.~\ref{sec:concl}. Throughout this paper we adopt the cosmology
of the original Millennium simulations ($H_0=73 ~\mathrm{km~s^{-1}}
~\mathrm{Mpc}^{-1}$, $\Omega_{\rm m}=0.25$, $\Omega_\Lambda=0.75$,
$n=1$). A discussion of the effect of cosmology on the satellite
properties presented in this paper is included in
Sec.~\ref{sec:tinf}.

\begin{table*}
\caption{Average halo virial radius $r_{\rm vir}$ (following G11),
  scale radius $r_s$ (following Zhao et al. 2009), inner radius
  $r_{\mathrm{inner}}$ and the $(g-r)$ colour cut separating blue from
  red satellites for the five primary stellar mass bins considered in
  our study. The final row gives the number of red and blue SDSS
  primaries in each of these bins.}
\begin{center}
\begin{tabular}{lrrrrr}\hline\hline
$\log M_*/M_\odot$ & \multicolumn{1}{c}{11.4-11.7} & \multicolumn{1}{c}{11.1-11.4} & \multicolumn{1}{c}{10.8-11.1} & \multicolumn{1}{c}{10.5-10.8} & \multicolumn{1}{c}{10.2-10.5} \\ \hline
$r_{\rm vir}$ [kpc]& 725 & 430 & 270 & 210 & 170 \\
$r_s$ [kpc] & 156.6 & 74.4 & 39.3 & 27.1 & 21.0 \\ 
$r_{\mathrm{inner}}$ [kpc] & 50 & 50 & 30 & 20 & 10 \\
$(g-r)_{\rm SDSS}$ & 0.840 & 0.830 & 0.820 & 0.811 & 0.801 \\
$(g-r)_{\rm mock}$ & 0.627 & 0.618 & 0.609 & 0.600 & 0.591 \\
$N_\mathrm{SDSS}$ [red, blue] & 1651, 35 & 6170, 731 & 8518, 4142 & 5453, 5953 & 1625,3764   \\
\hline
\label{tbl:primaries}
\end{tabular}
\end{center}
\end{table*}

\section{Data Selection}
\label{sec:data}

In this paper we use the same primary and satellite samples as in
Paper I. In the following, we briefly introduce the underlying
observational and simulation catalogues and the selection criteria
which define our samples, referring the reader to Paper I for further
details.

\subsection{Identification of primary and satellite galaxies}

We select isolated primary galaxies from the spectroscopic catalogue of 
the New York University Value Added Galaxy Catalogue (NYU-VAGC)
\footnote{http://sdss.physics.nyu.edu/vagc/}, which is built by \cite{Blanton2005} 
based on the Seventh Data Release of the Sloan Digital Sky Survey 
\citep[SDSS/DR7;][]{Abazajian2009}. Every galaxy with apparent (Petrosian) 
$r$-band magnitude brighter than $r=16.6$ is a primary candidate. To ensure 
isolation, it must fulfill two further conditions: it must $i)$ be at least 
one magnitude brighter than any companion within a projected radius of 
$r_p=0.5~{\rm Mpc}$ and a line-of-sight velocity difference 
$|\Delta{z}|<1000$~km/s, and $ii)$ be the brightest object within $r_p<1~{\rm Mpc}$ 
and $|\Delta{z}|<1000$~km/s. This returns 66,285 isolated primaries.

The SDSS spectroscopic sample is incomplete due to fibre-fibre collision.
For our selection criteria we expect $\sim$91.5\% completeness on average, 
varying with position on the sky
and worse in dense regions such as the centres 
of galaxy groups or clusters. To ensure that none of our primaries is falsely 
identified as isolated due to incompleteness in the spectroscopic survey, we 
look for further companions using the photometric SDSS catalogue. 
In practice, we will reject a primary candidate if it has a photometric 
companion satisfying
the position and magnitude cuts of i) and ii) which is 
absent from the spectroscopic catalogue but whose probability 
to have a redshift equal or less than the primary is larger than  $10\%$.
For this last step we use the photometric redshift distributions from \cite{Cunha2009}.
This reduces our sample of isolated primaries to 41,883 candidates. 
Lastly, we also consider survey boundaries to ensure that most of the 
companions of our primaries fall within the SDSS footprint. 
We use the spherical polygons provided on the NYU-VAGC website  
to quantify the survey boundaries and masked areas around bright stars. 
We remove from the above primary sample all candidates for which more than 20\% 
of a surrounding disk with $r_p<1 \rm Mpc$ lies outside the SDSS footprint. 
About 1.5\% galaxies are removed through this procedure, leading to our final 
primary sample with 41271 isolated systems.

\begin{center} \begin{figure*} 
\includegraphics[width=0.9\linewidth]{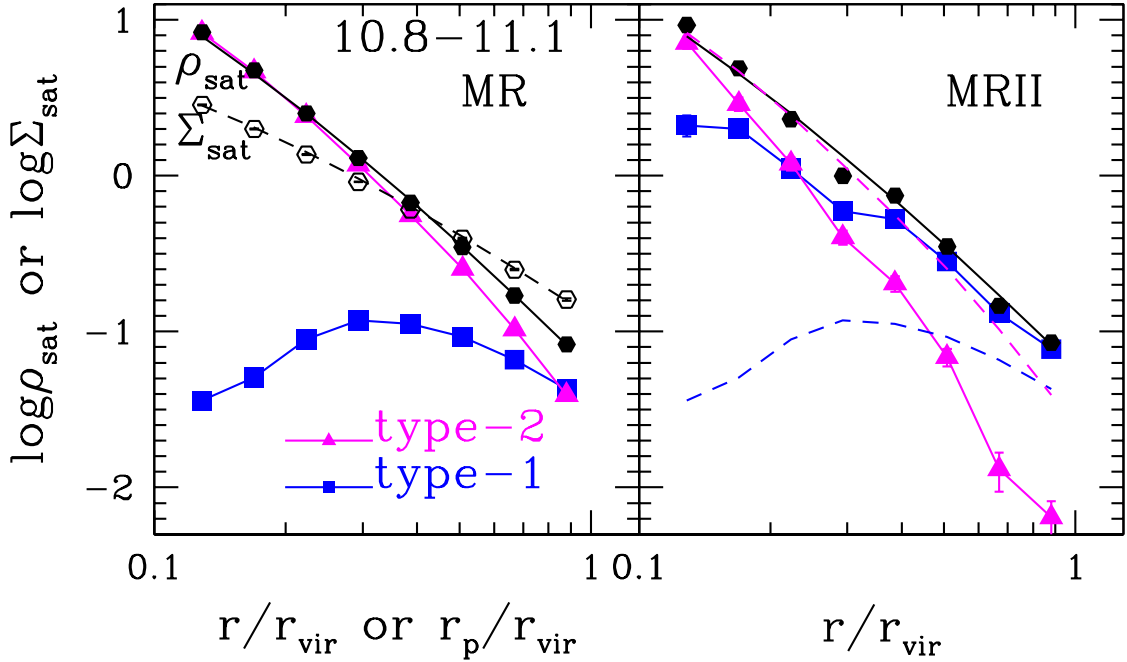}
\caption{The radial distribution of satellites with $\log
  M_*/M_\odot >9.0$ in semi-analytic catalogues based on the
  Millennium (left) and Millennium-II (right) simulations. Only
  isolated primaries in the stellar mass range $10.8<\log
  M_*/M_\odot<11.1$ are considered. Error bars are obtained from 
  the   scatter among 100 bootstrap re-sampled realizations of primaries, 
  and are smaller   than the marker size in most cases. Solid black 
  dots show the average 3D radial profile of satellites as a function of
  $r/r_{\rm vir}$. The distribution is well reproduced in both panels by
  an NFW profile with arbitrary normalization and the mean concentration 
  of the host dark matter halos (the black solid curve which is identical 
  in the two panels).  In each panel, the total profile is split into 
  contributions from satellite galaxies that still retain a dark matter 
  subhalo (blue solid squares) and from ``orphan'' galaxies that are 
  followed by the simulation even though their associated subhalo has been 
  tidally disrupted (magenta solid triangles). Orphans dominate the 
  satellite population in the Millennium but are important only in the 
  inner regions of Millennium-II halos, reflecting its 125 times better mass
  resolution. The data from the left panel are repeated as dashed blue
  and magenta curves in the right panel.  It is remarkable that the
  total satellite populations agree well between the two simulations
  despite the very different contribution of orphans in the two
  cases. Note also that orphans sill dominate in the inner regions
  even at MS-II resolution. Black empty circles in the left panel show
  the 2D radial distribution of satellites as a function of the
  projected distance $r_p/r_{\rm 200}$. This also follows well the
  projected NFW profile expected for the host dark matter halos (black
  dashed line). }
\label{fig:3Ddens}
\end{figure*}
\end{center}

Satellites are identified in the SDSS/DR8 photometric catalogue
\citep{Aihara2011} and are corrected statistically for background
contamination (see Paper I for details). We proceed as follows. For
each isolated primary we compute, as a function of projected distance
$r_p$, the number of objects with apparent magnitude $r$ and colour
$(g$-$r)$ in SDSS/DR8. For completeness, we only consider objects
brighter than $r=21$ (model magnitudes). For each bin in distance
$r_p$, we subtract the average number of galaxies in the $(r, g$-$r)$
bin expected in this area of the sky, as estimated from the survey as
a whole. The excess counts with respect to an homogeneous galaxy
background are assumed to be satellites physically associated with the
primary galaxy. Rest-frame colours and stellar masses can then be 
computed for these satellites by assigning them the redshift of the
primary.  Finally, results for different primaries can be averaged
after making completeness, volume and edge corrections as set out in
Paper I.

To aid the physical interpretation of these data, we use mock galaxy
catalogues generated from the Millennium and Millennium-II simulations
\citep{Springel2005a,Boylan-Kolchin2009}. The formation of galaxies is
simulated using the  semi-analytic model of \cite{Guo2011} (hereafter 
G11).  This is tuned to reproduce SDSS estimates of the mass 
and luminosity functions of low-redshift galaxies and also fits the
measured autocorrelations of SDSS galaxies at high stellar
mass. Autocorrelations of lower mass galaxies are significantly
overestimated on small scales ($r_p< 1.0~{\rm Mpc}$) but the
abundances of satellites around primaries of the mass considered in
this paper are in quite good agreement with direct measurements from
SDSS (see Paper I).  We create simulated galaxy catalogues by
projecting the simulation ``boxes'' in three orthogonal directions,
i.e.  parallel to their $x$, $y$ and $z$ axes. In each projection we
assign every galaxy a redshift based on its ``line-of-sight'' distance
and peculiar velocity. We can then apply similar isolation criteria as
in SDSS to identify a set of simulated primaries.

In addition, we analyze a sample of primary and satellite galaxies
derived from a ``light-cone'' mock catalogue which mimics
incompleteness due to fibre collisions as well the survey geometry of
SDSS/DR7 \citep{Henriques2012}\footnote{This catalogue is available at
  http://www.mpa-garching.mpg.de/millennium.}. This allows us to apply
{\it exactly} the same selection criteria as described above for the
SDSS sample and to compare with the simple projections used for our
main analysis (which provide better counting statistics). Results of
this comparison are shown in an Appendix.

\begin{center} \begin{figure*} 
\includegraphics[width=0.75\linewidth]{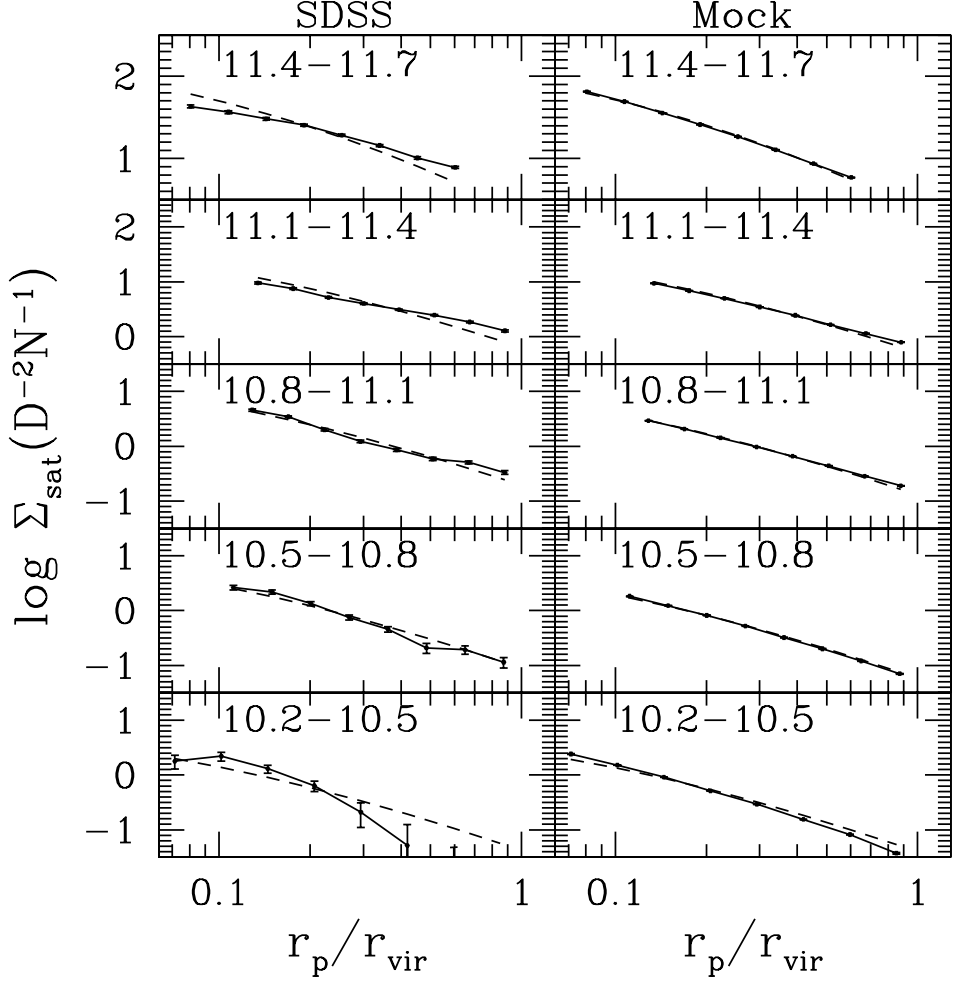}
\caption{Projected number density profiles for satellites brighter
  than $r$-band apparent magnitude $r=21$ and for primaries in 
  different stellar mass bins (the quoted numbers indicate the interval 
  in $\log M_*/M_\odot$).  The profiles correspond to the satellite 
  number counts per primary and per unit surface area  
  that we indicate in the y-axis 
  with units $N^{-1}$ and $D=r_p/r_\mathrm{vir}$,  respectively.
  Dashed lines indicate the expected distribution of dark matter
  around the hosts, computed by projecting the average NFW profile
  with mass following the $M_*-M_{\rm vir}$ relation from G11
  and concentration from Zhao et al. (2009).
  The shape of the distribution of SDSS satellites
  (left column) varies with primary stellar mass: for satellites of
  low-mass centrals ($ \log M_*/M_\odot<11.1$) the profile agrees
  well with the expected dark matter distribution. In
  contrast, satellites of the most massive primaries have a noticeably
  shallower distribution than predicted for the dark matter. Error
  bars are almost invisible and correspond to the scatter among 100
  bootstrap re-samplings of each primary sample. For comparison, the
  right column shows results of a similar analysis performed on our
  simulation catalogue. Satellites are here counted down to an
  $r$-band absolute magnitude corresponding to $r=21$ at the median
  redshift of the SDSS primaries in the corresponding left-hand
  panel. Note the excellent agreement in the absolute numbers of
  satellites in SDSS and in the simulation. The dependence of the
  shape of the satellite profiles on primary stellar mass is not,
  however, present in the simulation where satellites trace the
  underlying dark matter distribution regardless of primary stellar
  mass.}
\label{fig:densp_allprim}
\end{figure*}
\end{center}

\begin{center} \begin{figure*} 
\includegraphics[width=0.75\linewidth]{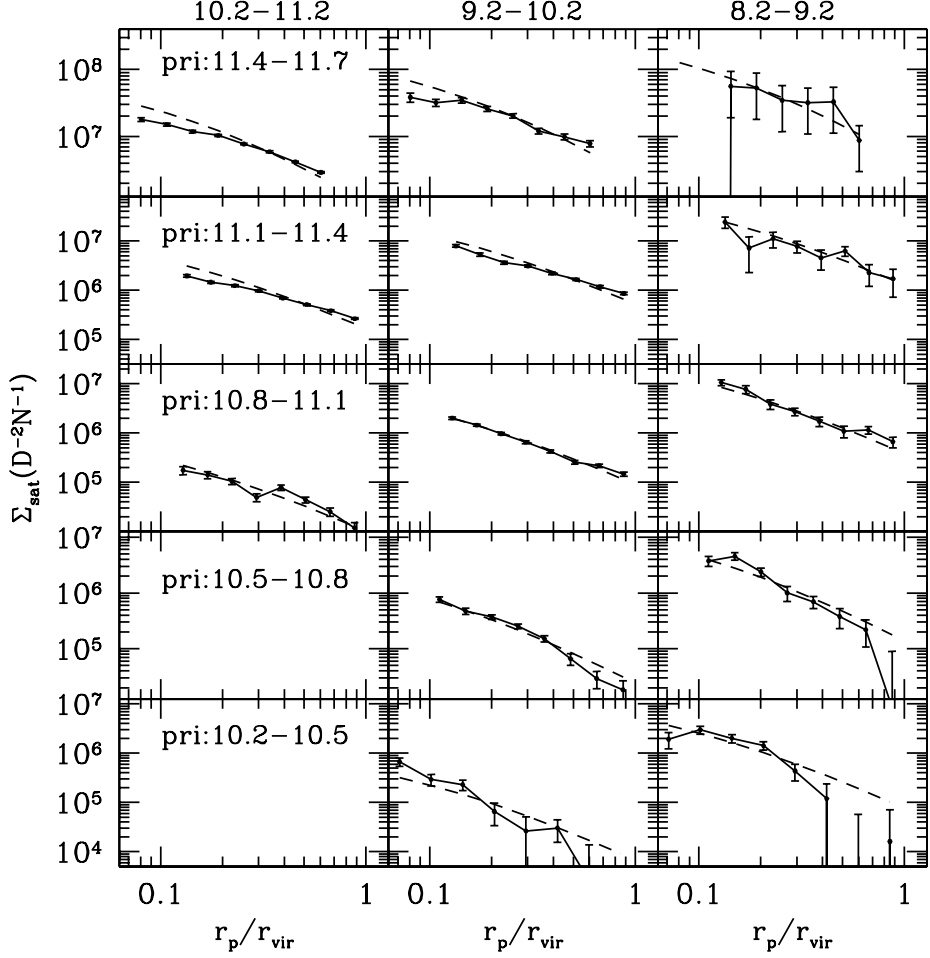}
\caption{ Projected number density profiles for satellites in SDSS
  split according to satellite stellar mass: $\log
  M_*/M_\odot =[10.2,11.2], [9.2,10.2]$ and $[8.2,9.2]$ from left to
  right. Samples divided by satellite stellar mass require the use of
  volume-limited samples, unlike Fig.~\ref{fig:densp_allprim}. 
  For a given primary mass bin, we find
  no strong dependence of satellite profiles with satellite stellar
  mass. Simulated satellites show a similar behaviour in the mock
  catalogue but are not shown here.}
\label{fig:densp_allprim_satmass}
\end{figure*}
\end{center}

\begin{center} \begin{figure*} 
\includegraphics[width=0.75\linewidth]{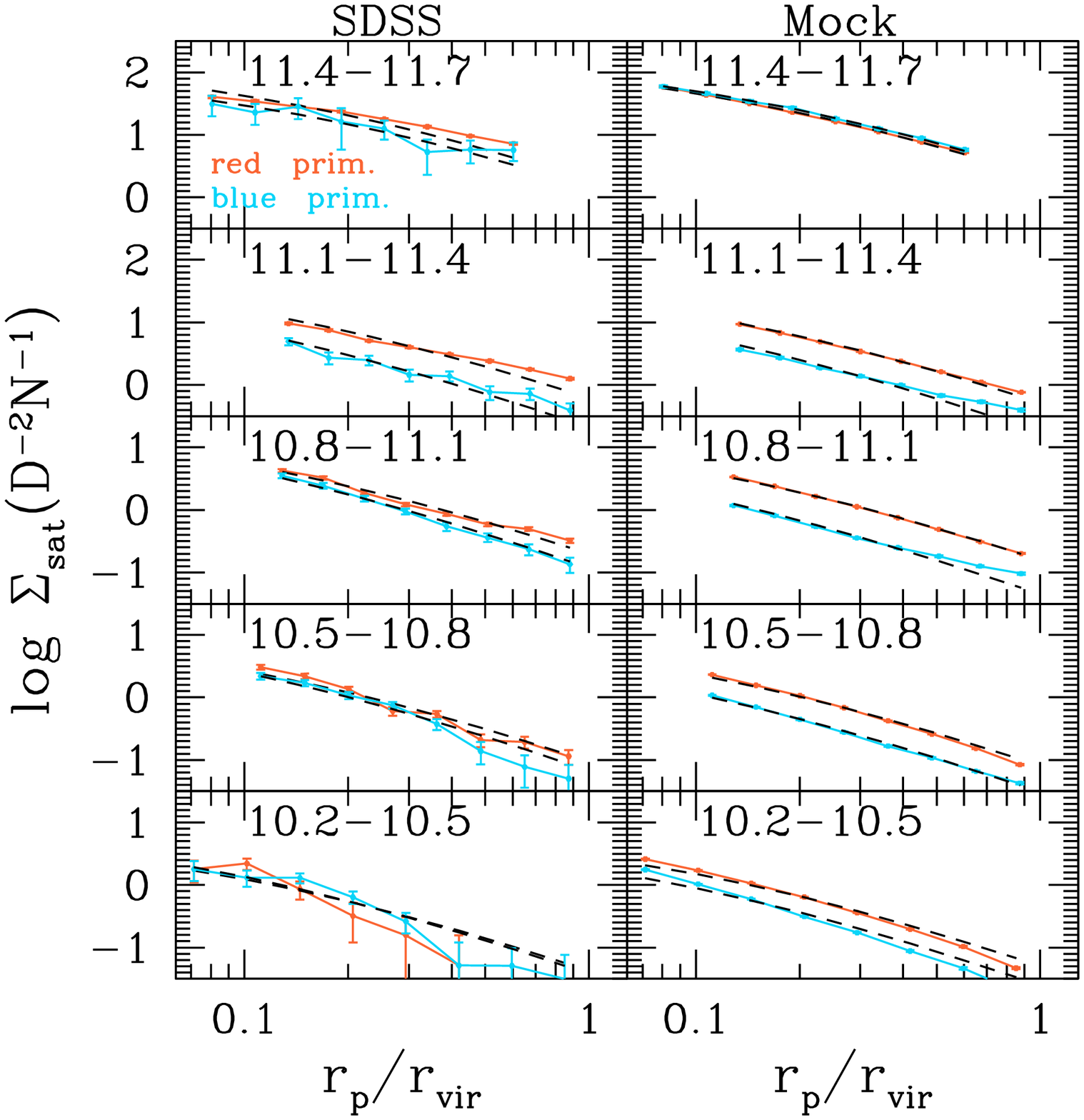}
\caption{As Fig.~\ref{fig:densp_allprim} but split according to
  primary colour; results for red primaries are shown in orange and
  for blue primaries in light-blue. As before, left and right columns
  show results from the SDSS and from the simulations,
  respectively. Black dashed curves are predicted NFW profiles for the
  host dark matter halos, re-normalized to fit the data. The overall
  shape of the satellite distribution does not depend on primary
  colour, except for the two most massive primary bins for SDSS, where
  the red primaries show slightly flatter profiles than the blue. Red
  primaries have a larger abundance of satellites at all radii, an
  effect that is more pronounced in the mock catalogue than in the
  SDSS. }
\label{fig:densp_primcolor}
\end{figure*}
\end{center}

\subsection{Satellite number density profiles}
\label{ssec:method}

The background subtraction method explained above provides a measure
of the cumulative number of satellites in fine grids of projected
distance in the range $r_p=[0,0.5] \rm Mpc$.  We group this data
according to primary mass and colour (Sec.~\ref{ssec:pricolor}) and to
satellite mass and colour (Sec.~\ref{ssec:satcolor}) in order to study
the average projected number density profile of satellites,
$\Sigma_{\rm sat}$,  defined as the mean number of satellites in 
some chosen
magnitude range per primary and per unit area as a function 
of
projected distance from the primary. Uncertainties are estimated 
from bootstrap re-samplings
of each primary sample. Following Paper I, 
we divide our primaries into 5 disjoint stellar mass bins: $\log
M_*/M_\odot=[10.2-10.5]$, $[10.5-10.8]$, $[10.8-11.1]$, $[11.1,11.4]$
and $[11.4-11.7]$.

For every primary stellar mass bins, we consider 8 equal-size
logarithmic radial bins that extend up to the average virial
radius\footnote{We define $r_{\rm vir}$ as the radius where the
  average enclosed density is $200$ times the critical value}, $r_{\rm
  vir}$, of the subsample (see Table 1). We estimate virial quantities
using the relation between mean halo mass and primary stellar mass for
all the isolated primaries in our semi-analytic catalogue. In SDSS the
luminosity of the primary galaxy affects the detectability of faint
satellites at small projected radii. We thus exclude the very central
regions and measure $\Sigma_{\rm sat}$ in the radial range $r_{\rm
  in}<r_p<r_{\rm vir}$, where $r_{\rm in}$ depends on primary stellar
mass as given in Table~\ref{tbl:primaries}. The average virial radius
for the most massive primary stellar mass bins is 725~kpc and
is thus larger than 500~kpc, the projected radius within which we
require that our primary galaxies be at least one magnitudes brighter
than any neighbour. This induces a feature in the satellite profile at
this radius, so below we present profiles for this largest primary
stellar mass bin only over the range $50~{\rm kpc}<r_p<500~{\rm kpc}$.

When considering the whole satellite population, we include as many
satellites as we can by going down to the flux limit, $r=21$. This
limit corresponds to different (intrinsic) satellite luminosities and
masses according to the redshift of the primary and the colour of the
satellite. Although by working with a flux-limited sample we are
including satellites of different masses for primaries at different
redshifts, this does not appreciably affect the global shape of the
satellite distribution we measure, because, as we show in
Sec.~\ref{ssec:primall} and \ref{ssec:satcolor}, there is only a very
weak dependence of $\Sigma_{\rm sat}$ on satellite stellar mass for
primaries in the mass range we consider.

When our analysis requires us to split satellites according to their
stellar mass and colour, we proceed as follows. Since satellites are
selected from the photometric catalogue, we assume they have the same
redshift as their primary galaxy, allowing us to convert observed
apparent magnitudes and colours into rest-frame quantities.  With
these, we estimate the stellar mass of the apparent companions
according to:
\begin{equation}
(M/L)_r=-1.0819^{0.1}(g-r)^2+4.1183^{0.1}(g-r)-0.7837.
\label{eq:mstr}
\end{equation}
\noindent  which is obtained through a fit to a flux-limited ($r < 17.6$) 
galaxy sample from the NYU-VAGC. Stellar masses in the sample were estimated 
by fitting stellar population synthesis models to the K-corrected galaxy
colours assuming a \citet{Chabrier2003} initial mass function as
in \citet{Blanton2007}. 

Notice that in contrast to the case when we take all satellites down
to the magnitude limit, when computing satellite number density
profiles as a function of satellite mass we need to account properly
for completeness limitations. In these cases we proceed as in Paper I:
a primary is allowed to contribute counts to a given bin in satellite
mass only if the K-corrected absolute luminosity corresponding to
$r=21$ for a galaxy at the redshift of the primary and lying on the
red envelope of the intrinsic colour distribution is fainter than the
lower luminosity limit of the bin.

Finally we note that our primary selection criteria result in a sample
of galaxies which are usually but not always the central galaxies of
their dark matter halos. In our simulated catalogue, the fraction of
primaries which are not the central object of their friends-of-friends
(FoF) dark matter halo is 0.1006, 0.1021, 0.0733, 0.0410 and 0.0156
for our five primary stellar mass bins (from most massive to least
massive). This fraction increases strongly with primary stellar mass
because of the concomitant increase in halo size (see Table~1). Note
that the great majority (80 to 90 percent) of these non-central
primaries actually lie outside the virial radius of the FoF halo of
which they are formally a satellite.

\section{Results}
\label{sec:results}

Throughout this paper we will make extensive use of comparisons
between results for the SDSS and for our simulated galaxy
populations. Before starting systematic presentation and
interpretation of the observed satellite profiles, we therefore use
our two large simulations to demonstrate that the relevant quantities
are {\it numerically} converged and that the observed, projected
radial profile of satellites is related as expected to the mean
three-dimensional distribution around the primary galaxies.  For
brevity, we focus on primaries with stellar mass in the range, $10.8
<\log M_*/M_\odot <11.1$ and satellites more massive than $\log
M_*/M_\odot>9$ , but we emphasize that we have checked explicitly
that similar results are found for other choices of primary and
satellite mass, as well as for samples split by primary or satellite
colour.

Fig.~\ref{fig:3Ddens} shows the 3D and 2D satellite number density
profiles in the Millennium and Millennium-II simulations. In both
panels, solid black dots show the three-dimensional number density of
satellites as a function of distance $r$ from their primary,
normalized to the average $r_{\rm vir}=270$~kpc for primaries of this
stellar mass (see Table~\ref{tbl:primaries}). These measurements agree
essentially perfectly given the counting statistics (there are 125
times fewer primaries in the MS-II) and are very well represented by
an NFW profile (black solid line) with concentration parameter $c=6.87$, 
the value expected for halos of virial mass $M_{\rm vir}=10^{12.5} \rm M_\odot$
\citep{Zhao2009}, which is the mean halo mass for simulated primaries
in this stellar mass bin.

\begin{center} \begin{figure*} 
\includegraphics[width=0.75\linewidth]{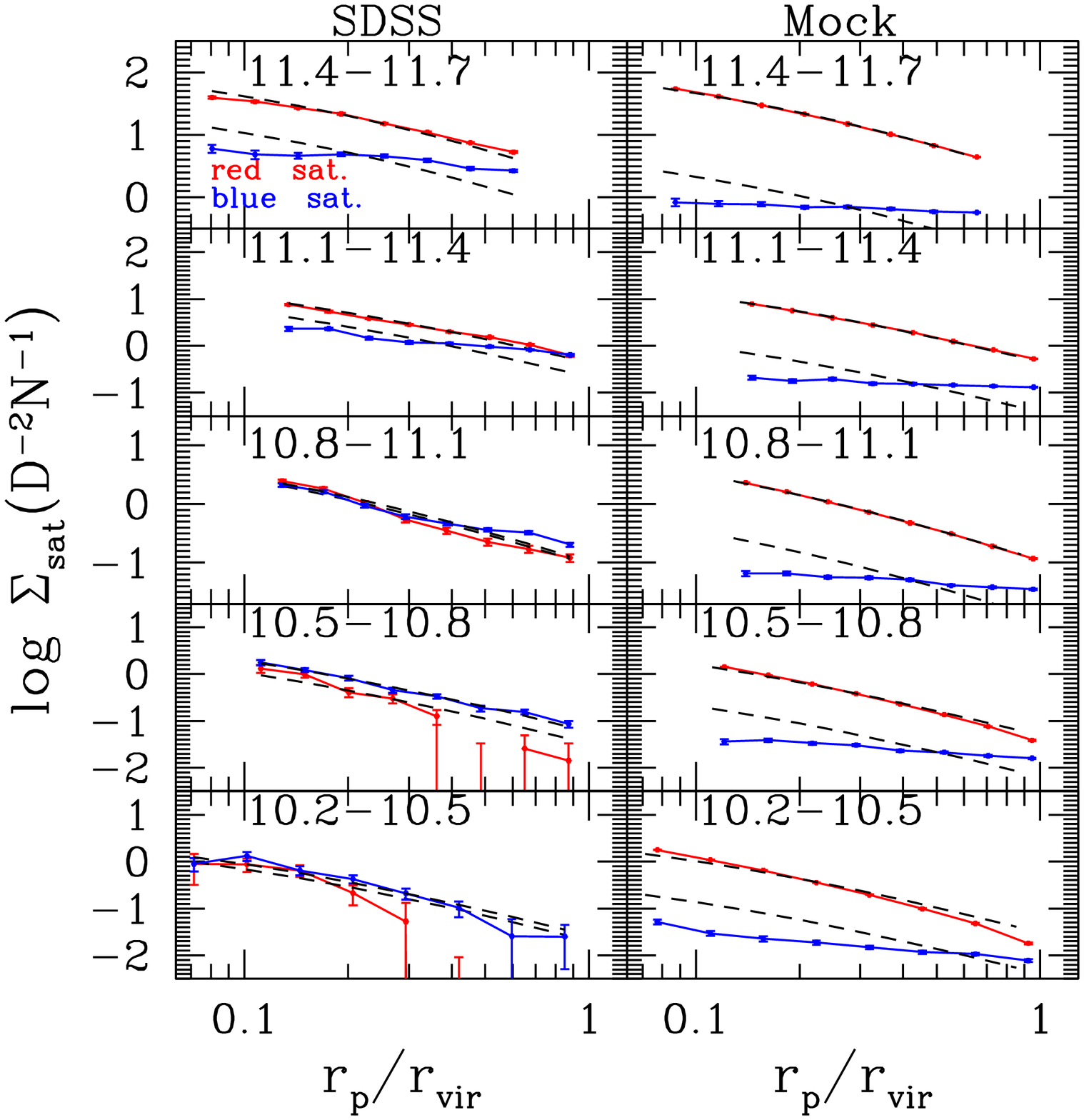}
\caption{Satellite number density profiles split according to
  satellite colour. Red and blue curves refer to red and blue
  satellites, respectively. Primaries are grouped by their stellar
  mass as before (rows). In the observed sample (left), the radial
  profile of red and blue satellites varies with primary stellar
  mass. For massive primaries, blue satellites have significantly
  shallower profiles than the red population and also than predicted
  for the dark matter (black dashed curves). However, for primaries
  with $\log M_*/M_\odot <11.1$, red and blue satellites have
  comparably steep profiles, which are similar to that expected for
  the dark matter.  Blue satellites dominate in number at all radii
  for these lower mass primaries.  The semi-analytic catalogue (right
  column) fails to reproduce many of these trends: blue satellites in
  the model always have a very shallow radial profile and are
  sub-dominant for all primary stellar masses. Environmental effects
  are apparently too strong in the semi-analytic model, particularly
  in low-mass halos. Red satellite profiles are similar to the dark
  matter predictions at all masses both in SDSS and in the
  simulations.}

\label{fig:densp_satcolor}
\end{figure*}
\end{center}

In the simulations there are two different kinds of satellites which
combine to give these number density profiles, those which have an
associated dark matter subhalo (type-1, blue solid squares in
Fig.~\ref{fig:3Ddens}) and those whose dark matter subhalo has fallen
below the resolution limit of the simulation (type-2, magenta solid 
triangles). The positions and velocities of the the latter ``orphan''
galaxies are set to the current values for the particles which were
the most bound at the centre of their subhalos at the last time these
were identified in the simulation.  Orphan galaxies are removed from
the galaxy catalogues when one of two conditions is fulfilled: either
the time since disruption of the subhalo exceeds the time estimated
for dynamical friction to cause a merger with the primary, or the
estimated tidal forces from the host exceed the binding energy of the
satellite so that it is disrupted (see G11 for further details.)

In the Millennium Simulation (the left panel of
Fig.~\ref{fig:3Ddens}), satellites with a dark matter subhalo are 
comparable in number to the orphans only near the virial radius.
Throughout the inner halo, satellite numbers are entirely dominated by
orphans. In the Millennium-II, however, (the right panel) orphans are
much less numerous and dominate the population only at the smallest
radii ($r < 0.2 r_{\rm vir}$).  To facilitate a direct comparison, we
repeat the type-1 and type-2 data from the left panel as dashed blue
and magenta lines in the right panel. The increase in mass resolution
in the MS-II increases the number of satellites with resolved subhalos
by factors between 3 and 30 in the inner halo and reduces the number
of orphans to compensate. Despite this, the total number of satellites
agrees very well and their profile is very similar to that of the dark
matter.  Note that even at the high resolution of MS-II it is
important to include the orphans to get a reliable and numerically
converged estimate of the number density profile. Similar conclusions
were drawn by G11 and by \citet{Guo2014} from the study of small scale
correlations and by \citet{Moster2010} from their abundance matching
work.

Fig.~\ref{fig:3Ddens} also explores projection effects in the
model. Black empty dots in the left panel show the average profile
obtained when systems are projected along their $x$, $y$ and $z$
axes. In practice, we count all apparent companions around our
primaries and correct statistically for unassociated objects that
happen to be projected near them, based on the mean surface number
density of galaxies times the area of each annular bin.  This
projected number density profile, $\Sigma_{\rm sat}$, is also shown as
a function of projected distance $r_p$ normalized to $r_{\rm vir}$,
and again is very well represented by an NFW fit to the (projected)
dark matter distribution (dashed black line).  Thus, the simulations
suggest that satellite profiles should closely parallel the mean dark
matter distributions around isolated galaxies, a result that could be
checked directly using galaxy-galaxy lensing.

In what follows we will study projected number density profiles of
satellites around isolated SDSS galaxies, comparing directly with
simulation results. We have checked that numerical convergence between
the two simulations is as good as in Fig.~\ref{fig:3Ddens} for all the
other plots we show, and hence, unless otherwise stated, we show only
results based on the Millennium Simulation, since these have better
counting statistics.

\begin{center}
\begin{figure*}
\includegraphics[width=0.86\linewidth]{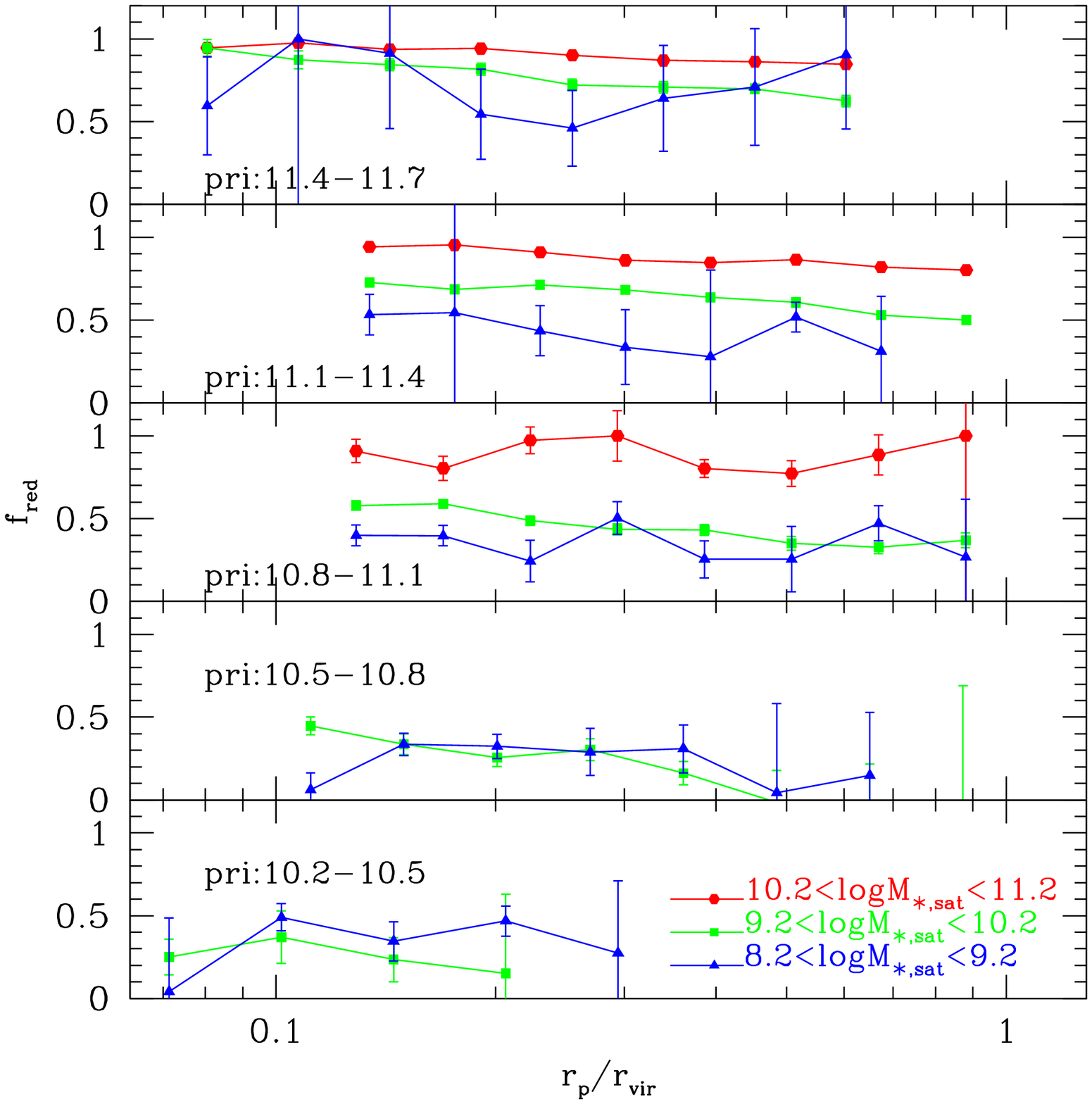}
\caption{The fraction of red satellites, $f_{\rm red}$, as a function
  of projected distance from isolated SDSS primary galaxies. Each
  panel corresponds to primaries in a different range of $\log
  M_*/M_\odot$, as indicated. Results for satellites in three different
  stellar mass ranges are shown as curves of different colour. Red
  satellites are more common around high-mass primaries and also at
  smaller distances, although the trends with radius are rather
  weak. Error bars correspond to 100 bootstrap re-samplings of the
  primary samples.}
\label{fig:redfrac}
\end{figure*}
\end{center}

\subsection{Satellite number density profiles: dependence on
  primary stellar mass}
\label{ssec:primall}

The thick black lines in the left column of
Fig.~\ref{fig:densp_allprim} show the mean projected number density of
satellites around isolated SDSS primaries, $\Sigma_{\rm sat}$, 
as a function of projected separation ($r_p$), normalized by mean inferred 
virial radius ($D=r_p/r_\mathrm{vir}$). Each row corresponds to primaries in 
a different stellar mass range, as indicated by the listed values of 
$\log M_*/M_\odot$. Satellite number density profiles $\Sigma_{\rm sat}$ are 
computed by summing the satellite counts (after background correction) in 
each radial bin, and then dividing by the number of contributing primaries 
per mass bin (see Table~\ref{tbl:primaries}). The error bars correspond to 
the dispersion among 100 bootstrap re-samplings of each primary sample.
We maximize the statistics by including all satellites down to the apparent 
magnitude limit $m_r=21$ of the SDSS photometric catalogue. Because of the 
large number of satellites included (including background galaxies, there
are about $\sim 7,000,000$ photometric companions projected within
500~kpc of these primaries), the error bars are almost invisible in
most cases.

{\it We find that the shape of satellite profiles around our isolated 
SDSS primaries depends significantly on primary stellar mass, with a 
shallower radial distribution around high-mass primaries than around 
primaries with $\log M_*/M_\odot < 11.1$.} This is most clearly 
seen by comparing with the expected mean dark matter profiles, which
we indicate as dashed lines in each plot.  These are computed by 
using the stellar mass of each primary to estimate its halo mass 
according to the mean $M_*$-$M_{200}$ relation in the semi-analytic 
catalogue of G11. We then compute the average halo mass in each bin 
and use the concentration-mass relation of \citet{Zhao2009} to get 
the mean projected NFW profile expected for the dark matter. Finally, 
we re-adjust the normalization to obtain a good fit to the amplitude 
of the simulation results in the right panel of each row (the dashed 
curves are identical in each pair of panels). We have explicitly 
checked that stacking the DM particles directly around the simulated 
galaxy sample gives almost identical results.

Massive primaries with $\log M_*/M_\odot > 11.1$ show a
satellite profile that is inconsistent with the predicted dark matter
profile.  In contrast, the satellite distribution agrees 
well with the predicted dark matter distribution around lower mass
primaries. In the lowest mass bin (the bottom left panel) the
satellite profile declines more steeply than the predicted dark matter
profile at large radii. However, our tests indicate that at these
radii the satellite counts around low-mass primaries become sensitive
to background subtraction and are artificially steepened because our
isolation criteria cause a slight suppression of the number of {\it
  background} galaxies around our primaries compared to randomly
selected points on the sky.  We show in the Appendix that the measured 
satellite profiles in a light-cone mock catalogue decline more steeply 
in the lowest mass bin than those measured directly from the projected 
simulation box. Such uncertainties in background subtraction do not affect 
our profile measurements at smaller radii or around more massive galaxies, 
because of the higher mean densities expected there. 

The right column of Fig.~\ref{fig:densp_allprim} compares our SDSS
results with analogous results for satellites surrounding isolated
primaries in our mock catalogue based on the G11 simulation. For these
plots, simulated satellites are counted down to an $r$-band absolute
magnitude corresponding to $r=21$ at the median redshift of the SDSS
primaries in the corresponding left panel. As noted above, the dashed
curve in each panel is an NFW profile representing the mean halo mass
distribution of the simulation primaries with its normalization
adjusted to fit the satellite counts, and is identical to the curve
overplotted on the SDSS data in the corresponding left panel. The
excellent agreement in normalization in each pair of panels thus
repeats the result of Paper I, that the G11 simulation does a good job
in reproducing the observed abundance of satellites as a function of
primary mass. However, in the simulation, satellites follow the dark
matter distribution for primaries of all stellar masses\footnote{We
  have checked that this is true independent of satellite mass,
  although, for brevity, we do not show the figure.}.  This is
inconsistent with the shallower slope found for massive SDSS
primaries.

This disagreement is puzzling in view of the good match
between model and SDSS found by G11 for the small scale
autocorrelations of massive galaxies, and the significantly poorer
agreement in shape found for the autocorrelations of lower mass
galaxies (see Fig. 20 in G11). Apparently, although massive primaries
have the right spatial distribution in the mock catalogue, their
satellites are somewhat too concentrated to small radii in comparison
to SDSS. We show in the Appendix that more relaxed isolation 
criteria result in flatter
satellite profiles in the outer regions
(since satellites associated to other nearby primaries are now allowed 
to contribute), but the change is only significant for low/intermediate 
mass primaries\footnote{Although in the Appendix we only show results for 
the SDSS sample, we have checked that a similar effect is obtained in the 
simulated mock catalogue when using the more relaxed isolation criteria}. 
The overall shallower profile of satellites around massive primaries in 
SDSS seems robust to changes in isolation criteria and its physical 
explanation remains unclear. More efficient tidal disruption associated 
to massive primaries could provide a viable explanation for the inner 
shallow slopes. More definitive conclusions will require a better 
treatment of this process in the models, together, perhaps,
with studies of the intergalactic light \citep[e.g.][]{Conroy2007,Contini2014,Presotto2014} 
systems.

{\it We conclude that SDSS satellite galaxies are, in the mean, distributed 
in the same way as the dark matter around isolated galaxies like the Milky 
Way or M31, but are somewhat less centrally concentrated around more  
massive systems.}  The latter feature is not well reproduced in the model, 
where satellites follow the dark matter profile for primaries in all mass bins.

We investigate this further in Fig.~\ref{fig:densp_allprim_satmass} by
splitting our SDSS satellites according to their stellar mass. As
explained in Sec.~\ref{ssec:method}, the completeness corrections
needed to build unbiased, satellite-mass-limited subsamples reduce the
primary sample by $\sim 50\%$ overall. As a result, the uncertainties
in Fig.~\ref{fig:densp_allprim_satmass} are larger than in the
previous figure, especially for the less massive satellites where the
volume surveyed is considerably smaller.
Fig.~\ref{fig:densp_allprim_satmass} shows that the shallower profiles 
around massive primaries are quite pronounced for massive satellites 
(left and middle columns) but are not significantly detected for the 
lowest mass satellites (right column) where the data are much 
noisier. Nevertheless, variations of satellite profile with satellite 
mass are weak or undetected for all primary masses.

\subsection{Satellite number density profiles: dependence on primary colour}
\label{ssec:pricolor}

We explore the dependence of satellite radial profile on primary
colour in Fig.~\ref{fig:densp_primcolor}. We split primaries according
to their $(g-r)$ colour at a value that varies weakly with primary
stellar mass (see Table~\ref{tbl:primaries}).  These cuts are the same as
used in Paper I (see Fig. 2 there) and reflect the position of the trough
between the red and blue peaks in the colour distribution of isolated
primaries in each $M_*$ bin. Because the colour
distributions of observed and simulated primaries differ slightly, we
list separately in Table~\ref{tbl:primaries} the colour cuts used in
the SDSS and the mock galaxy samples, $(g-r)_{\rm SDSS}$ and
$(g-r)_{\rm mock}$, respectively.

When comparing satellite abundances, it is important to account for
the different redshift distributions of red and blue primaries. Red
primaries are located at systematically lower redshift, resulting in
an offset in the intrinsic luminosities of satellites between the two
primary populations if satellites are counted to a fixed apparent
magnitude limit (e.g. $r=21$ as above, see Sec.~\ref{sec:data}).  To
make the comparison unbiased, we count satellites down to $r=21$
around blue primaries, but only down to $r=$20.76, 20.75, 20.61, 20.66
and 20.44 around red primaries in the most massive to least massive
stellar mass bins quoted in table~\ref{tbl:primaries}. We compute
these limits from the difference in distance modulus between the
median redshifts of red and blue primaries (0.24, 0.25, 0.39, 0.34 and
0.56) which was then subtracted from $r=21$.

We show in the left column of Fig.~\ref{fig:densp_primcolor} the
average satellite number density profiles for red and blue SDSS
primaries (solid orange and light-blue curves, respectively). The {\it
  normalizations} of the curves in Fig.~\ref{fig:densp_primcolor}
depend on primary colour in all stellar mass bins. Red primaries have
systematically more satellites at every radius, particularly for the
most massive bins ($\log M_*/M_\odot>10.8$). This is
consistent with the results in Paper I, where we reported a larger
total number of satellites around red than around blue primaries at
fixed stellar mass.  Fig.~\ref{fig:densp_primcolor} shows that these
``excess'' companions are distributed more or less evenly across the
full radial range.  An excess of satellites around red primaries at
fixed {\it luminosity} had previously been reported in a number of
other studies \citep{Sales2007c,GuoQuan2012,Wang2012}.

Black dashed curves in Fig.~\ref{fig:densp_primcolor} show the
predicted dark matter profiles, obtained as in
Fig.~\ref{fig:densp_allprim} and Fig.~\ref{fig:densp_allprim_satmass}.
Since the halo mass differs between red and blue primaries of similar
stellar mass, we estimate dark matter profiles for red and blue
primaries separately using the $M_*$-$M_{200}$ relations for red and blue
primaries in our semi-analytic catalogue.  These have been
re-normalized to fit the amplitude of the measured satellite
profiles. For primaries more massive than $\log M_*/M_\odot=11.1$,
satellites around red primaries have significantly shallower profiles
than predicted for the dark matter. The profiles around blue
primaries are noisier and steeper, but still show some tendency to be
shallower than predicted for the dark matter.

In the right column of Fig.~\ref{fig:densp_primcolor} we show
predictions from our simulated catalogue for comparison with the
observed profiles. Overall, the qualitative agreement
between models and data, is quite good although the difference in the
normalization between red and blue primaries is more pronounced in the
models than in the SDSS for primaries less massive than $\log
M_*/M_\odot=11.1$. Note that the shapes of the dark matter
profiles in the left column of Fig.~\ref{fig:densp_primcolor} are
based on those of the the simulated primaries in the right column, and
so may be biased by this difference in colour dependence.

There appears to be an excess of companions at large radii around blue
primaries with $10.8<\log M_*/M_\odot<11.4$, especially in the right
column of Fig.~\ref{fig:densp_primcolor}. We have checked and found
that this is mostly due to the small fraction of our primary sample
which, despite passing all our isolation criteria, are nevertheless
actually satellite galaxies in massive groups or clusters. The excess
counts reflect a contribution from fainter galaxies within these
groups/clusters. A similar bump in the SDSS data may have been
weakened due to the over-subtraction of background counts at large
distances (see the Appendix for more details): it still can be seen
around primaries with $11.1<\log M_*/M_\odot<11.4$.  The excess count
is substantially weakened if we additionally require all our simulated
primaries to be the central galaxies of their FoF halos.  We have
checked and find that this bump in the satellite distribution
parallels a similar excess at large radii in the mean profiles for the 
dark matter stacked around these hosts, in good agreement with our
previous conclusion that simulated satellites trace the underlying dark 
matter in considerable detail (see also the Appendix).

\subsection{Satellite number density profiles: dependence on satellite colour}
\label{ssec:satcolor}

\begin{table}
\caption{$(g-r)$ colour cuts for the three satellite 
  stellar mass bins considered in our study.}
\begin{center}
\begin{tabular}{lrrrrr}\hline\hline
           & \multicolumn{1}{c}{10.2-11.2} & \multicolumn{1}{c}{9.2-10.2} & \multicolumn{1}{c}{8.2-9.2} \\ \hline
$(g-r)_{\rm SDSS}$ & 0.796 & 0.731 & 0.666  \\
$(g-r)_{\rm mock}$ & 0.606 & 0.526 & 0.446  \\
\hline
\label{tbl:satellites}
\end{tabular}
\end{center}
\end{table}

We explore the radial distribution of satellites according to their
colour in Fig.~\ref{fig:densp_satcolor}. The left column shows
results for the SDSS sample. We consider all primaries in a given
$M_*$ bin, splitting the satellites into red and blue subsamples
(solid red and blue curves, respectively).  For this, we use a colour
boundary that depends on satellite stellar mass and corresponds to the
trough between the blue and red peaks of the satellite colour
distributions.  Table~\ref{tbl:satellites} lists the satellite stellar
mass bins and colour cuts used in our study.

The left column of Fig.~\ref{fig:densp_satcolor} shows that the
dependence of number density profile on satellite colour is
complex. For massive primaries ($\log M_*/M_\odot>11.1$), red
satellites have steeper profiles than blue ones, and only the former
are an approximate tracer of the expected dark matter distribution
(black dashed line). Blue satellites have a shallow profile and are
sub-dominant at almost all radii. This behaviour changes, however, for
lower primary mass.  For $\log M_*/M_\odot<11.1$, the red and blue
satellite populations have similar profiles and both are similar to
the expected halo dark matter distribution. At these primary masses,
the dominant satellite population is blue.

These results suggest that environmental effects are a strong function
of primary stellar mass in our SDSS sample, or equivalently, of host halo
mass. Satellites orbiting massive primaries tend to be red,
particularly if they are close to the primary. As a result, there is a
deficiency of blue objects relative to red in the inner regions. On
the other hand, for primaries with $M_*<10^{11} M_\odot$,
environmental effects are sufficiently weak that satellites can
continue to form stars even if they are close to their primaries. The
blue population thus maintains the steep profile characteristic of the
dark matter and dominates by number at all radii.

This result can be seen more clearly in Fig.~\ref{fig:redfrac}, which
shows the fraction of red satellites as a function of radius in each
of our primary stellar mass bins. Because galaxy colours depend
intrinsically on stellar mass, we show $f_{\rm red}$ separately for
satellites in three different stellar mass ranges: 
$\log M_*/M_\odot=[8.2-9.2], [9.2,10.2] $ and $[10.2-11.2]$ in blue,
green and red, respectively. In agreement with
Fig.~\ref{fig:densp_satcolor}, $f_{\rm red}$ is larger than 0.5 only
for the most massive primary bins; most satellites remain blue for
primaries with $\log M_*/M_\odot<10.8$.  Notice that although $f_{\rm
  red}$ decreases with radius, the dependence is weak, indicating that
environmental effects on satellite colour depend relatively little on
distance to the primary.

We compare these SDSS results with profiles from our simulation
catalogue in the right column of Fig.~\ref{fig:densp_satcolor}.
Despite the relatively good agreement seen in previous figures, the
models do very poorly at reproducing profiles split by satellite
colour. This is primarily because, as already noted in Paper I, the
fraction of satellites around low-mass primaries which are blue is
substantially too low, the discrepancy reaching an order of magnitude
at the lowest masses. The profiles for the few blue satellites which
remain in the simulation are much shallower than the dark matter
profiles, regardless of primary stellar mass.  Clearly, satellite
colours -- and presumably related properties such as specific star
formation rates, morphologies, gas fractions -- are poorly
reproduced by the model, indicating that it is substantially
overestimating the effects of environment on these properties of
satellites. A similar conclusion was reached by \cite{GuoQuan2013} for
a different simulation of similar type to the one we analyze here.

\begin{figure}
\includegraphics[width=84mm]{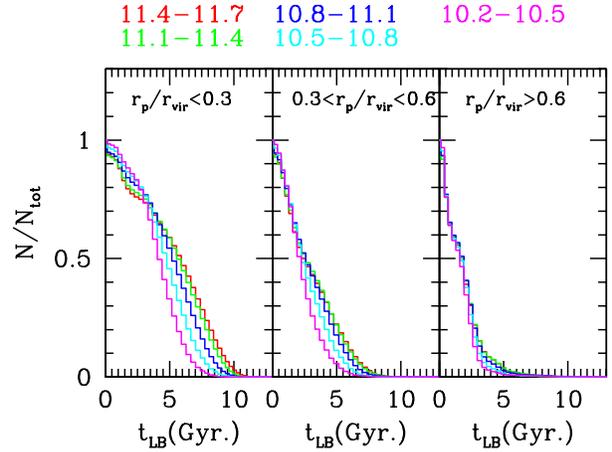}
\caption{Cumulative distribution of (look-back) infall times $t_{LB}$
  for satellites with $\log M_{\ast}/M_\odot>9$ in the semi-analytic
  galaxy catalogue.  Colored curves distinguish our five primary
  stellar mass bins as labeled.  We show three different ranges of
  projected radius: $r_p<0.3 r_{\rm vir}$ (left), $0.3<r_p/r_{\rm
    vir}<0.6$ (middle) and $r_p>0.6 r_{\rm vir}$ (right). Satellites
  in the inner regions typically fell in earlier than those in the
  outskirts by $\sim 2-3$ Gyr. This is consistent with the inside-out
  assembly of the halo and its satellite population, which is thus
  detected also in projection.  About half of the satellites close to
  primaries with $M_*^{\rm pri}<11.1$ fell in more than 5 Gyr
  ago. Since Fig.~\ref{fig:densp_satcolor} shows that the satellite
  population around low-mass SDSS primaries is dominated by blue
  objects, this suggests that the timescale for quenching star
  formation in satellites is at least $\sim 5$ Gyrs for such
  primaries.  }
\label{fig:tinf}
\end{figure}

\begin{figure}
\epsfig{figure=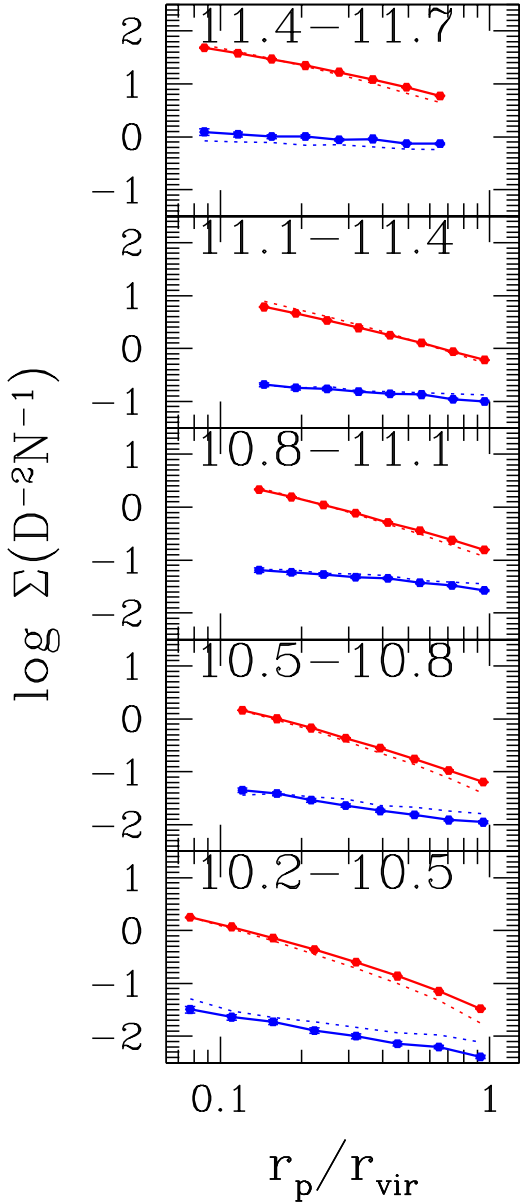,width=0.42\textwidth}
\caption{The effects of background cosmology on satellite profiles.
  Dotted/solid lines show the same physical model but for two
  different sets of cosmological parameters, WMAP-1 (G11) and WMAP-7
  (G13), respectively. The impact of cosmology is apparently quite
  small.  }
\label{fig:cosmology}
\end{figure}

\subsection{Satellite number density profiles: comparison with previous work}
\label{ssec:comparison}

A number of recent studies have examined the distribution of satellites 
in hybrid samples from spectroscopic + photometric catalogues using methods 
similar to our own \citep{Lares2011,GuoQuan2012,Nierenberg2011,Nierenberg2012, 
Tal2012,Jiang2012,GuoHong2014}. Most of them agree with us in finding the 
abundance of satellites to depend strongly on primary stellar mass (see also 
PaperI). However, in several cases the detailed trends we find with primary 
and satellite properties do not all agree with those published previously. 
For instance, the very weak dependence of the shape of satellite profiles 
on primary colour and satellite mass agrees well with Nierenberg et al.(2012), 
\citet{Wang2011} and \citet{Jiang2012} but is contrary to some of those in 
\citet{Tal2012}, \citet{Watson2012} and \citet{GuoHong2014} who find bright 
satellites to be more radially concentrated. Our results are also in partial 
disagreement with \citet{GuoQuan2012}, who found the radial distribution of 
bright satellites to be less radially concentrated.  
 
Some of these discrepancies can be explained by differences in sample
definition. Our selection of isolated primaries discards all galaxy
systems where the difference of $r$-band magnitude between the central
object and the satellites is smaller than 1. This makes the comparison
of our results with those found in groups and clusters more difficult
to interpret. However, comparisons with \citet{GuoQuan2012} are
more interesting since their analysis uses similar selection criteria
to our own and is also applied to objects in the SDSS/DR7 and DR8
catalogues.

The approaches to quantifying satellite radial distributions differ
between our work and \citet{GuoQuan2012}: we use abundance matching
arguments to infer the expected mean distribution of dark matter
around our primary samples, and we compare this with the mean number
density profiles we find for satellites, whereas Guo et al. fit NFW
profiles directly to their estimated satellite distributions without
reference to expectations for the host dark matter halos. The shapes
of the number density profiles they measure for satellites of
different luminosity agree for $r_p>0.12r_\mathrm{vir}$, but differ at
smaller radii (see their Figure 6).  The variations in the inner 
profiles are thus not totally unexpected, given the different choices and 
cuts applied to each sample. Here, we assume a fixed projected radius cut
for all primaries in a given mass bin (see Table~\ref{tbl:primaries}) whereas 
\citet{GuoQuan2012} deal with the
inner regions by excluding annuli 
that are within 1.5 times the Petrosian radius of the primary galaxy. 
This is
typically a smaller radius than the cut we impose. For example, 
the luminosity range of primaries in \citet{GuoQuan2012} corresponds roughly  
to the second most massive primary stellar mass bin in our analysis, for 
which we use $r_{\rm in}=50~{\rm kpc} \approx 0.12r_\mathrm{vir}$ (see 
Table~\ref{tbl:primaries}), while the mean inner cut of \citet{GuoQuan2012} 
is $r_{\rm in}=$23kpc and differs from galaxy to galaxy. On scales smaller 
than $0.1r_\mathrm{vir}$ systematics caused by proximity to the primary image 
are argued to be important by both \cite{Tal2012} and \cite{Watson2012}, 
whereas no photometric corrections are made by \citet{GuoQuan2012}, which 
perhaps accounts for the different conclusions about inner profiles in these 
three studies. We chose such a more conservative inner radius cut because we 
have tested that satellite profiles at these radii are sensitive to photometric 
systematics (see the Appendix of paper I for more details). We exclude these 
radii from our analysis specifically to avoid the need to correct for such 
effects.

\section{Treatment of environmental effects in semi-analytical catalogues}
\label{sec:tinf}

The serious discrepancy between satellite colours in the SDSS and in
our semi-analytic catalogue clearly reflects an overestimation of
environmental effects in the simulation. Once a simulated galaxy
becomes a satellite, i.e.  crosses the virial radius of a larger
system, its external and internal gas reservoirs are reduced through
tidal and ram-pressure stripping, with no further replenishment
through cosmological infall. The reduction in fuel for star formation
then causes satellites to form fewer stars and to redden compared to
similar objects in the field. Studies by \citet{Guo2011} of
colour-dependent autocorrelation functions in their model already
revealed excess clustering of red galaxies on scales below $\sim 1$
Mpc. Our analysis of satellite properties allows a cleaner
interpretation, tracing back the origin of the problem to an incorrect
treatment of star formation for objects that orbit within a larger
host halo.  This problem with the G11 model was already clearly
identified in the study of satellite colours in galaxy groups and
clusters by \citet{Weinmann2010}.  These authors showed that although
discrepancies were clearly smaller in G11 than in the earlier model
studied by \citet{Weinmann2010}, environmental effects were still too
strong in lower mass systems.

We use the distribution of time since infall for simulated satellites
to gain intuition on the appropriate timescale for suppression of star
formation once a galaxy becomes a satellite. Fig.~\ref{fig:tinf} shows
the cumulative distribution of time since infall for satellites more
massive than $\log M_{\ast}/M_\odot=9$. Time since infall $t_{\rm LB}$
is here defined as the time since the satellite last crossed the
virial radius of its host halo. We group the satellites in our mock
sample into three bins of projected radius normalized to the virial
radius: from left to right these are $r_p<0.3 r_{\rm vir}$,
$0.3<r_p/r_{\rm vir}<0.6$ and $r_p>0.6 r_{\rm vir}$.  This allows us
to see the dependence of infall time on distance to the host
\citep{Gao2004a}, or equivalently on binding energy \citep{Rocha2012}.
Different colours indicate results for primaries in the five stellar
mass bins defined above. Note that we only consider ``true''
satellites, defined as objects with 3D positions within the virial
radius of each host halo, when we make this plot, although we bin as a
function of projected radius to enable more direct comparison with
observed systems.

\begin{figure}
\epsfig{figure=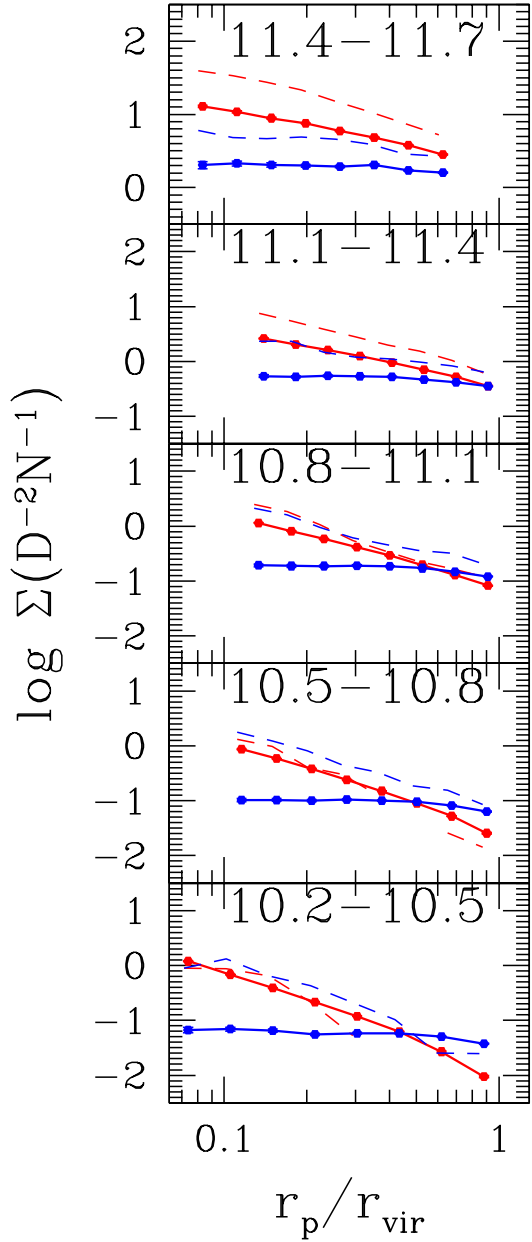,width=0.42\textwidth}%
\caption{Prediction from an updated semi-analytic model (from
  Henriques et al. (2014, in prep.)) which, relative to the G11 model,
  has delayed reincorporation of gas ejected by supernovae, a lower
  star formation threshold, and no ram-pressure stripping in galaxy
  groups. The fraction of blue satellites (thick solid lines) is
  increased significantly, but the profiles remain flatter than those
  observed in SDSS (the thin dashed curves).  }
\label{fig:model_test2}
\end{figure}

As expected, satellites in the inner regions typically have a longer
time since infall than those at large radii, consistent with
inside-out growth of the satellite population. Interestingly, we find
that the time since infall is also typically longer in more massive
halos. This is surprising, because dark matter concentrations and
formation redshifts are typically lower for more massive halos. The
trend we see is in part due to the fact that our isolated primaries
are usually the central galaxies of fossil groups (galaxy groups with
a large magnitude difference between the first and second brightest
galaxies) and these assemble earlier than typical groups of their
mass.  The second brightest companions in our mock catalogue are  
typically 1.9, 2.4, 2.8, 2.8 and 2.5 magnitudes fainter in r-band than 
their primaries in the most- to least- massive stellar mass bin, 
respectively.

Fig.~\ref{fig:tinf} shows that around isolated galaxies similar in
mass to the Milky Way (the blue curves), half of all satellites that
today have $r_p<0.3 r_{\rm vir}$ first fell within the virial radius
of the primary more than 5~Gyr ago, whereas for satellites with
$r_p>0.6 r_{\rm vir}$ the median time since infall is $\sim 2~{\rm
  Gyr}$.  Fig.~\ref{fig:densp_satcolor} and \ref{fig:redfrac} show
that most observed satellites of primaries with $\log M_*/M_\odot <11.1$ 
are blue, even at small radii, so the timescale for shutting-off star 
formation apparently needs to be at least 5~Gyr in such systems.
For comparison, in the G11 model analyzed here, the mean time since 
infall for blue satellites of primaries with
$\log M_*/M_\odot =11.1$ is only $\sim 0.9~{\rm Gyr}$.


If satellites in the G11 model redden too quickly compared to
observation, could overly early collapse and halo assembly (due to an
overly large $\sigma_8$ value) be responsible?  We show in
Fig.~\ref{fig:cosmology} the effect of cosmology on satellite profiles
split by colour. Thick solid lines correspond to the
\citet[][]{Guo2013} model (G13), which is the same as G11 except that
it corrects the cosmological parameters to be consistent with 
WMAP7. The differences from the original G11 model (the thin dotted
lines) are small, indicating that the assumed background cosmology
has little effect on the properties of satellites and thus cannot
explain the discrepancy between the semi-analytic model and the SDSS
data.

We take a closer look at the impact of different physical assumptions
on the predicted profiles of red and blue satellites by analyzing
predictions from an updated model specifically targeted at improving
the treatment of satellite and low-mass galaxies. This model
(Henriques et al. 2014, in prep) is based on G13 and adopts most of
its physical prescriptions.  An important modification is an increase
in the timescale for reincorporation of material ejected to large
radius by supernova explosions, particularly in low-mass systems and
early times. This follows \citet{Henriques2013} and extends star
formation to later times in low-mass galaxies. It has a large impact
on the properties of central galaxies but the strong environmental
effects in G11/G13 prevent it from substantially increasing the number
of blue satellites.

To address this problem the updated model also reduces the threshold
for star formation and removes the effects of ram-pressure stripping
in galaxy groups. (Tidal stripping is assumed to act in groups of all
masses, whereas ram-pressure effects are eliminated for group masses
$M_{\rm{vir}}<10^{14}\Msun$.) These changes cause satellites in
low-mass groups to retain their extended gas reservoirs for longer,
and to convert more of their cold gas into stars. Together these
modifications ensure that the bulk of low-mass satellites remains blue
at later times. This model also updates the cosmological parameters to
be consistent with PLANCK results, and modifies the scaling of the AGN
radio-mode feedback to improve the properties of high-mass
galaxies. These last two changes have negligible impact on the
properties analyzed here.

Fig.~\ref{fig:model_test2} shows the distribution of red and blue
satellites in the updated model (thick solid lines). Observations from
SDSS are indicated by thin dashed lines.  The new recipes produce a
significant increase in the number of blue satellites around primaries
of all masses. As a result, the normalization of the predicted blue
population gets closer to observations. However, the profile shape for
this blue population is only similar to that observed for the two most
massive primary bins. Profiles are still significantly shallower than 
observed for low-mass primaries.

These results, together with an extensive series of tests in which we
have arbitrarily suppressed individual environmental effects (not
shown here for brevity) seem to indicate that straightforward
modification of standard environmental effects cannot lead to a
satellite population which remains predominantly blue at small
separations.  A possible solution might involve the {\it enhancement}
of star formation in satellites near halo centre, perhaps as a result
of tidal effects induced by the primary.  There is some direct
observational evidence that star formation can indeed be triggered by
close tidal interactions \citep[e.g., ][]{Lambas2003,Li2008,Ellison2008}, 
and starbursts which may be
tidally induced have been detected in the nuclei of S0 galaxies in the
Fornax and Virgo clusters by \cite{Johnston2012,Johnston2013}, showing
that star formation can take place in the bulge of these galaxies even
as their disks redden with time. Furthermore, \cite{Ebeling2014} used
high-resolution HST images to uncover evidence of shock-induced star
formation in cluster galaxies undergoing ram-pressure stripping. Such
environmental enhancement of star formation may counteract quenching
processes to keep the fraction of star-forming satellites high even at
small distances from lower mass primaries.


\section{Conclusion and Discussion}
\label{sec:concl}

We study the mean number density profiles of satellites around
isolated primary galaxies selected from the spectroscopic catalogue of
SDSS/DR7. We select satellites from the full photometric catalogue of
SDSS/DR8, correcting statistically for contamination by unrelated
foreground and background galaxies. Our sample contains about 41,000
isolated primaries with $\sim 7,000,000$ photometric companions
(including background) projected within 500~kpc. We explore the
dependence of these profiles on the stellar mass and colour of both
primaries and satellites. Our results can be summarized as follows:

\begin{itemize}

\item The radial distribution of satellites depends on primary stellar
  mass. Satellites around massive primaries $\rm
  \log M_*/M_\odot>11.1$ have slightly shallower profiles than are
  predicted for the dark matter in their host halos, whereas for less
  massive primaries $10.2 <\log M_*/M_\odot< 11.1$ satellites
  follow quite closely the predicted dark matter profiles.

\item We find the shape of satellite number density profiles to depend 
  at most weakly on satellite stellar mass.

\item Red primaries have more satellites than blue primaries of the
  same stellar mass, at least for $\log M_*/M_\odot> 10.8$.

\item Observed satellite number density profiles depend on satellite
  colour and behave differently for high- and low-mass primaries. For
  primaries with $\log M_*/M_\odot<11.1$, the blue and red
  populations have profiles of similar shape, consistent in both cases
  with that predicted for the dark matter distribution. Blue
  satellites dominate at all radii. Around more massive primaries
  ($\log M_*/M_\odot>11.1$) the blue population has a shallower
  profile and is subdominant at all radii.
\end{itemize}

We compare these observational results with satellite samples selected
from the galaxy population simulation of \citet{Guo2011}.  The number 
density profiles of the whole satellite population always parallel the
dark matter profile of the host halo, regardless of primary mass; this
disagrees with the SDSS result for massive primaries.  This may
reflect the need for more efficient tidal disruption of satellites in
the model. Satellite colours remain the
most important challenge to these theoretical models, however. In the
model the fraction of blue satellites is too small, particularly
around low-mass primaries, and the few remaining blue satellites have
an almost flat radial profile, in clear disagreement with the
observations.

Given that observed satellites of low-mass primaries are predominantly
blue at all projected radii, the distributions of time since infall
that we find for simulated satellites imply that real satellites can
remain actively star-forming as much as 5 Gyr after they have fallen
into their current host halo, even when their orbit takes them into
its inner regions. This seems qualitatively consistent with earlier
work reporting that the decline of star formation in satellites occurs
over extended periods of time \citep[e.g.][]{WangLi2007,Weinmann2009,
Wetzel2013,Trinh2013,Wheeler2014}.
The significantly shorter timescales implied by our model ($\sim 0.9$
Gyr) are responsible for its overabundance of red satellites. This
indicates that the environmental suppression of star formation is
overestimated by the model, particularly around low-mass primaries,
and perhaps that the environmental stimulation of star formation needs
to be included.

Indeed, from a series of experiments with differing treatments of
environmental processes, we find that although the combined effect of
suppressing ram-pressure stripping in low-mass halos
($M_\mathrm{vir}<10^{14}M_\odot$) and decreasing the density
threshold for star formation can bring the overall blue fraction into
agreement with SDSS, the {\it shape} of the blue satellite profile
remains much flatter than observed. The fact that SDSS satellites are
still predominantly blue even a few tens of kpc from low-mass
primaries, suggests that processes which enhance star formation during
close encounters need to be introduced into the models. Progress in
this area will require a better understanding of tidally or shock-induced
star formation, as well as observational studies which resolve the
structure of star-forming regions in typical satellite galaxies.

\section*{Acknowledgements}
Wenting Wang is partially supported by NSFC (11121062, 
10878001, 11033006, 11003035), by the CAS/SAFEA 
International Partnership Program for Creative Research
 Teams (KJCX2-YW-T23) and by the Science and Technology 
Facilities Council (ST/F001166/1). LS was supported in part 
by the Marie Curie RTN CosmoComp. BH and SW were supported 
by ERC Advanced Grant 246797 GALFORMOD. Wenting Wang is 
grateful for useful discussions with Yipeng Jing about 
the underlying methodology, sample selection and survey 
geometry.


\bibliography{master}

\begin{thebibliography}{}

\bibitem[\protect\citeauthoryear{{Abazajian}, {Adelman-McCarthy},
  {Ag{\"u}eros}, {Allam}, {Allende Prieto}, {An}, {Anderson}, {Anderson},
  {Annis}, {Bahcall} \& et al.}{{Abazajian} et~al.}{2009}]{Abazajian2009}
{Abazajian} K.~N.,  {Adelman-McCarthy} J.~K.,  {Ag{\"u}eros} M.~A.,  {Allam}
  S.~S.,  {Allende Prieto} C.,  {An} D.,  {Anderson} K.~S.~J.,  {Anderson}
  S.~F.,  {Annis} J.,  {Bahcall} N.~A.,    et al. 2009, \apjs, 182, 543

\bibitem[\protect\citeauthoryear{{Aihara} et~al.,}{{Aihara}
  et~al.}{2011}]{Aihara2011}
{Aihara} H.,  et~al., 2011, \apjs, 193, 29

\bibitem[\protect\citeauthoryear{{Blanton} \& {Roweis}}{{Blanton} \&
  {Roweis}}{2007}]{Blanton2007}
{Blanton} M.~R.,  {Roweis} S.,  2007, \aj, 133, 734

\bibitem[\protect\citeauthoryear{{Blanton}, {Schlegel}, {Strauss}, {Brinkmann},
  {Finkbeiner}, {Fukugita}, {Gunn}, {Hogg}, {Ivezi{\'c}}, {Knapp}, {Lupton},
  {Munn}, {Schneider}, {Tegmark} \& {Zehavi}}{{Blanton}
  et~al.}{2005}]{Blanton2005}
{Blanton} M.~R.,  {Schlegel} D.~J.,  {Strauss} M.~A.,  {Brinkmann} J.,
  {Finkbeiner} D.,  {Fukugita} M.,  {Gunn} J.~E.,  {Hogg} D.~W.,  {Ivezi{\'c}}
  {\v Z}.,  {Knapp} G.~R.,  {Lupton} R.~H.,  {Munn} J.~A.,  {Schneider} D.~P.,
  {Tegmark} M.,    {Zehavi} I.,  2005, \aj, 129, 2562

\bibitem[\protect\citeauthoryear{{Bower}, {Benson}, {Malbon}, {Helly}, {Frenk},
  {Baugh}, {Cole} \& {Lacey}}{{Bower} et~al.}{2006}]{Bower2006}
{Bower} R.~G.,  {Benson} A.~J.,  {Malbon} R.,  {Helly} J.~C.,  {Frenk} C.~S.,
  {Baugh} C.~M.,  {Cole} S.,    {Lacey} C.~G.,  2006, \mnras, 370, 645

\bibitem[\protect\citeauthoryear{{Boylan-Kolchin}, {Springel}, {White},
  {Jenkins} \& {Lemson}}{{Boylan-Kolchin} et~al.}{2009}]{Boylan-Kolchin2009}
{Boylan-Kolchin} M.,  {Springel} V.,  {White} S.~D.~M.,  {Jenkins} A.,
  {Lemson} G.,  2009, \mnras, 398, 1150

\bibitem[\protect\citeauthoryear{{Budzynski}, {Koposov}, {McCarthy}, {McGee} \&
  {Belokurov}}{{Budzynski} et~al.}{2012}]{Budzynski2012}
{Budzynski} J.~M.,  {Koposov} S.~E.,  {McCarthy} I.~G.,  {McGee} S.~L.,
  {Belokurov} V.,  2012, \mnras, 423, 104

\bibitem[\protect\citeauthoryear{{Chabrier}}{{Chabrier}}{2003}]{Chabrier2003}
{Chabrier} G.,  2003, \apjl, 586, L133

\bibitem[\protect\citeauthoryear{{Chen}}{{Chen}}{2008}]{Chen2008}
{Chen} J.,  2008, \aap, 484, 347

\bibitem[\protect\citeauthoryear{{Chen}, {Kravtsov}, {Prada}, {Sheldon},
  {Klypin}, {Blanton}, {Brinkmann} \& {Thakar}}{{Chen} et~al.}{2006}]{Chen2006}
{Chen} J.,  {Kravtsov} A.~V.,  {Prada} F.,  {Sheldon} E.~S.,  {Klypin} A.~A.,
  {Blanton} M.~R.,  {Brinkmann} J.,    {Thakar} A.~R.,  2006, \apj, 647, 86

\bibitem[\protect\citeauthoryear{{Coil}, {Newman}, {Croton}, {Cooper}, {Davis},
  {Faber}, {Gerke}, {Koo}, {Padmanabhan}, {Wechsler} \& {Weiner}}{{Coil}
  et~al.}{2008}]{Coil2008}
{Coil} A.~L.,  {Newman} J.~A.,  {Croton} D.,  {Cooper} M.~C.,  {Davis} M.,
  {Faber} S.~M.,  {Gerke} B.~F.,  {Koo} D.~C.,  {Padmanabhan} N.,  {Wechsler}
  R.~H.,    {Weiner} B.~J.,  2008, \apj, 672, 153

\bibitem[\protect\citeauthoryear{{Colless}, {Dalton}, {Maddox}, {Sutherland},
  {Norberg}, {Cole}, {Bland-Hawthorn} \& {Bridges}}{{Colless}
  et~al.}{2001}]{Colless2001}
{Colless} M.,  {Dalton} G.,  {Maddox} S.,  {Sutherland} W.,  {Norberg} P.,
  {Cole} S.,  {Bland-Hawthorn} J.,    {Bridges} 2001, \mnras, 328, 1039

\bibitem[\protect\citeauthoryear{{Conroy}, {Wechsler} \& {Kravtsov}}{{Conroy}
  et~al.}{2007}]{Conroy2007}
{Conroy} C.,  {Wechsler} R.~H.,    {Kravtsov} A.~V.,  2007, \apj, 668, 826

\bibitem[\protect\citeauthoryear{{Contini}, {De Lucia}, {Villalobos} \&
  {Borgani}}{{Contini} et~al.}{2014}]{Contini2014}
{Contini} E.,  {De Lucia} G.,  {Villalobos} {\'A}.,    {Borgani} S.,  2014,
  \mnras, 437, 3787

\bibitem[\protect\citeauthoryear{{Croton}, {Springel}, {White}, {De Lucia},
  {Frenk}, {Gao}, {Jenkins}, {Kauffmann}, {Navarro} \& {Yoshida}}{{Croton}
  et~al.}{2006}]{Croton2006}
{Croton} D.~J.,  {Springel} V.,  {White} S.~D.~M.,  {De Lucia} G.,  {Frenk}
  C.~S.,  {Gao} L.,  {Jenkins} A.,  {Kauffmann} G.,  {Navarro} J.~F.,
  {Yoshida} N.,  2006, \mnras, 365, 11

\bibitem[\protect\citeauthoryear{{Cunha}, {Lima}, {Oyaizu}, {Frieman} \&
  {Lin}}{{Cunha} et~al.}{2009}]{Cunha2009}
{Cunha} C.~E.,  {Lima} M.,  {Oyaizu} H.,  {Frieman} J.,    {Lin} H.,  2009,
  \mnras, 396, 2379

\bibitem[\protect\citeauthoryear{{Ebeling}, {Stephenson} \& {Edge}}{{Ebeling}
  et~al.}{2014}]{Ebeling2014}
{Ebeling} H.,  {Stephenson} L.~N.,    {Edge} A.~C.,  2014, \apjl, 781, L40

\bibitem[\protect\citeauthoryear{{Ellison}, {Patton}, {Simard} \&
  {McConnachie}}{{Ellison} et~al.}{2008}]{Ellison2008}
{Ellison} S.~L.,  {Patton} D.~R.,  {Simard} L.,    {McConnachie} A.~W.,  2008,
  \aj, 135, 1877

\bibitem[\protect\citeauthoryear{{Font}, {Bower}, {McCarthy}, {Benson},
  {Frenk}, {Helly}, {Lacey}, {Baugh} \& {Cole}}{{Font} et~al.}{2008}]{Font2008}
{Font} A.~S.,  {Bower} R.~G.,  {McCarthy} I.~G.,  {Benson} A.~J.,  {Frenk}
  C.~S.,  {Helly} J.~C.,  {Lacey} C.~G.,  {Baugh} C.~M.,    {Cole} S.,  2008,
  \mnras, 389, 1619

\bibitem[\protect\citeauthoryear{{Gao}, {De Lucia}, {White} \& {Jenkins}}{{Gao}
  et~al.}{2004}]{Gao2004b}
{Gao} L.,  {De Lucia} G.,  {White} S.~D.~M.,    {Jenkins} A.,  2004, \mnras,
  352, L1

\bibitem[\protect\citeauthoryear{{Gao}, {Frenk}, {Boylan-Kolchin}, {Jenkins},
  {Springel} \& {White}}{{Gao} et~al.}{2011}]{Gao2011}
{Gao} L.,  {Frenk} C.~S.,  {Boylan-Kolchin} M.,  {Jenkins} A.,  {Springel} V.,
    {White} S.~D.~M.,  2011, \mnras, 410, 2309

\bibitem[\protect\citeauthoryear{{Gao}, {Navarro}, {Cole}, {Frenk}, {White},
  {Springel}, {Jenkins} \& {Neto}}{{Gao} et~al.}{2008}]{Gao2008}
{Gao} L.,  {Navarro} J.~F.,  {Cole} S.,  {Frenk} C.~S.,  {White} S.~D.~M.,
  {Springel} V.,  {Jenkins} A.,    {Neto} A.~F.,  2008, \mnras, 387, 536

\bibitem[\protect\citeauthoryear{{Gao}, {White}, {Jenkins}, {Stoehr} \&
  {Springel}}{{Gao} et~al.}{2004}]{Gao2004a}
{Gao} L.,  {White} S.~D.~M.,  {Jenkins} A.,  {Stoehr} F.,    {Springel} V.,
  2004, \mnras, 355, 819

\bibitem[\protect\citeauthoryear{{Guo}, {Zheng}, {Zehavi}, {Xu}, {Eisenstein},
  {Weinberg}, {Bahcall}, {Berlind}, {Comparat}, {McBride}, {Ross}, {Schneider},
  {Skibba}, {Swanson}, {Tinker}, {Tojeiro} \& {Wake}}{{Guo}
  et~al.}{2014}]{GuoHong2014}
{Guo} H.,  {Zheng} Z.,  {Zehavi} I.,  {Xu} H.,  {Eisenstein} D.~J.,  {Weinberg}
  D.~H.,  {Bahcall} N.~A.,  {Berlind} A.~A.,  {Comparat} J.,  {McBride} C.~K.,
  {Ross} A.~J.,  {Schneider} D.~P.,  {Skibba} R.~A.,  {Swanson} M.~E.~C.,
  {Tinker} J.~L.,  {Tojeiro} R.,    {Wake} D.~A.,  2014, ArXiv e-prints

\bibitem[\protect\citeauthoryear{{Guo}, {Cole}, {Eke} \& {Frenk}}{{Guo}
  et~al.}{2012}]{GuoQuan2012}
{Guo} Q.,  {Cole} S.,  {Eke} V.,    {Frenk} C.,  2012, \mnras, 427, 428

\bibitem[\protect\citeauthoryear{{Guo}, {Cole}, {Eke}, {Frenk} \&
  {Helly}}{{Guo} et~al.}{2013}]{GuoQuan2013}
{Guo} Q.,  {Cole} S.,  {Eke} V.,  {Frenk} C.,    {Helly} J.,  2013, ArXiv
  e-prints

\bibitem[\protect\citeauthoryear{{Guo} \& {White}}{{Guo} \&
  {White}}{2014}]{Guo2014}
{Guo} Q.,  {White} S.,  2014, \mnras, 437, 3228

\bibitem[\protect\citeauthoryear{{Guo}, {White}, {Angulo}, {Henriques},
  {Lemson}, {Boylan-Kolchin}, {Thomas} \& {Short}}{{Guo}
  et~al.}{2013}]{Guo2013}
{Guo} Q.,  {White} S.,  {Angulo} R.~E.,  {Henriques} B.,  {Lemson} G.,
  {Boylan-Kolchin} M.,  {Thomas} P.,    {Short} C.,  2013, MNRAS, 428, 1351

\bibitem[\protect\citeauthoryear{{Guo}, {White}, {Boylan-Kolchin}, {De Lucia},
  {Kauffmann}, {Lemson}, {Li}, {Springel} \& {Weinmann}}{{Guo}
  et~al.}{2011}]{Guo2011}
{Guo} Q.,  {White} S.,  {Boylan-Kolchin} M.,  {De Lucia} G.,  {Kauffmann} G.,
  {Lemson} G.,  {Li} C.,  {Springel} V.,    {Weinmann} S.,  2011, \mnras, 413,
  101

\bibitem[\protect\citeauthoryear{{Henriques}, {White}, {Lemson}, {Thomas},
  {Guo}, {Marleau} \& {Overzier}}{{Henriques} et~al.}{2012}]{Henriques2012}
{Henriques} B.~M.~B.,  {White} S.~D.~M.,  {Lemson} G.,  {Thomas} P.~A.,  {Guo}
  Q.,  {Marleau} G.-D.,    {Overzier} R.~A.,  2012, \mnras, 421, 2904

\bibitem[\protect\citeauthoryear{{Henriques}, {White}, {Thomas}, {Angulo},
  {Guo}, {Lemson} \& {Springel}}{{Henriques} et~al.}{2013}]{Henriques2013}
{Henriques} B.~M.~B.,  {White} S.~D.~M.,  {Thomas} P.~A.,  {Angulo} R.~E.,
  {Guo} Q.,  {Lemson} G.,    {Springel} V.,  2013, \mnras, 431, 3373

\bibitem[\protect\citeauthoryear{{Holmberg}}{{Holmberg}}{1969}]{Holmberg1969}
{Holmberg} E.,  1969, Arkiv for Astronomi, 5, 305

\bibitem[\protect\citeauthoryear{{Jiang}, {Jing} \& {Li}}{{Jiang}
  et~al.}{2012}]{Jiang2012}
{Jiang} C.~Y.,  {Jing} Y.~P.,    {Li} C.,  2012, \apj, 760, 16

\bibitem[\protect\citeauthoryear{{Johnston}, {Arag{\'o}n-Salamanca},
  {Merrifield} \& {Bedregal}}{{Johnston} et~al.}{2012}]{Johnston2012}
{Johnston} E.~J.,  {Arag{\'o}n-Salamanca} A.,  {Merrifield} M.~R.,
  {Bedregal} A.~G.,  2012, \mnras, 422, 2590

\bibitem[\protect\citeauthoryear{{Johnston}, {Aragon-Salamanca}, {Merrifield}
  \& {Bedregal}}{{Johnston} et~al.}{2013}]{Johnston2013}
{Johnston} E.~J.,  {Aragon-Salamanca} A.,  {Merrifield} M.~R.,    {Bedregal}
  A.~G.,  2013, ArXiv e-prints

\bibitem[\protect\citeauthoryear{{Kravtsov}, {Gnedin} \& {Klypin}}{{Kravtsov}
  et~al.}{2004}]{Kravtsov2004}
{Kravtsov} A.~V.,  {Gnedin} O.~Y.,    {Klypin} A.~A.,  2004, \apj, 609, 482

\bibitem[\protect\citeauthoryear{{Lambas}, {Tissera}, {Alonso} \&
  {Coldwell}}{{Lambas} et~al.}{2003}]{Lambas2003}
{Lambas} D.~G.,  {Tissera} P.~B.,  {Alonso} M.~S.,    {Coldwell} G.,  2003,
  \mnras, 346, 1189

\bibitem[\protect\citeauthoryear{{Lares}, {Lambas} \&
  {Dom{\'{\i}}nguez}}{{Lares} et~al.}{2011}]{Lares2011}
{Lares} M.,  {Lambas} D.~G.,    {Dom{\'{\i}}nguez} M.~J.,  2011, \aj, 142, 13

\bibitem[\protect\citeauthoryear{{Li}, {Kauffmann}, {Heckman}, {Jing} \&
  {White}}{{Li} et~al.}{2008}]{Li2008}
{Li} C.,  {Kauffmann} G.,  {Heckman} T.~M.,  {Jing} Y.~P.,    {White} S.~D.~M.,
   2008, \mnras, 385, 1903

\bibitem[\protect\citeauthoryear{{Li}, {Kauffmann}, {Jing}, {White},
  {B{\"o}rner} \& {Cheng}}{{Li} et~al.}{2006}]{LiCheng2006}
{Li} C.,  {Kauffmann} G.,  {Jing} Y.~P.,  {White} S.~D.~M.,  {B{\"o}rner} G.,
   {Cheng} F.~Z.,  2006, \mnras, 368, 21

\bibitem[\protect\citeauthoryear{{Lorrimer}, {Frenk}, {Smith}, {White} \&
  {Zaritsky}}{{Lorrimer} et~al.}{1994}]{Lorrimer1994}
{Lorrimer} S.~J.,  {Frenk} C.~S.,  {Smith} R.~M.,  {White} S.~D.~M.,
  {Zaritsky} D.,  1994, \mnras, 269, 696

\bibitem[\protect\citeauthoryear{{Mandelbaum}, {Hirata}, {Broderick}, {Seljak}
  \& {Brinkmann}}{{Mandelbaum} et~al.}{2006}]{Mandelbaum2006}
{Mandelbaum} R.,  {Hirata} C.~M.,  {Broderick} T.,  {Seljak} U.,    {Brinkmann}
  J.,  2006, \mnras, 370, 1008

\bibitem[\protect\citeauthoryear{{Moore}, {Ghigna}, {Governato}, {Lake},
  {Quinn}, {Stadel} \& {Tozzi}}{{Moore} et~al.}{1999}]{Moore1999}
{Moore} B.,  {Ghigna} S.,  {Governato} F.,  {Lake} G.,  {Quinn} T.,  {Stadel}
  J.,    {Tozzi} P.,  1999, \apjl, 524, L19

\bibitem[\protect\citeauthoryear{{Moster}, {Somerville}, {Maulbetsch}, {van den
  Bosch}, {Macci{\`o}}, {Naab} \& {Oser}}{{Moster} et~al.}{2010}]{Moster2010}
{Moster} B.~P.,  {Somerville} R.~S.,  {Maulbetsch} C.,  {van den Bosch} F.~C.,
  {Macci{\`o}} A.~V.,  {Naab} T.,    {Oser} L.,  2010, \apj, 710, 903

\bibitem[\protect\citeauthoryear{{Nagai} \& {Kravtsov}}{{Nagai} \&
  {Kravtsov}}{2005}]{Nagai2005}
{Nagai} D.,  {Kravtsov} A.~V.,  2005, \apj, 618, 557

\bibitem[\protect\citeauthoryear{{Navarro}, {Frenk} \& {White}}{{Navarro}
  et~al.}{1996}]{Navarro1996}
{Navarro} J.~F.,  {Frenk} C.~S.,    {White} S.~D.~M.,  1996, \apj, 462, 563

\bibitem[\protect\citeauthoryear{{Navarro}, {Frenk} \& {White}}{{Navarro}
  et~al.}{1997}]{Navarro1997}
{Navarro} J.~F.,  {Frenk} C.~S.,    {White} S.~D.~M.,  1997, \apj, 490, 493

\bibitem[\protect\citeauthoryear{{Nierenberg}, {Auger}, {Treu}, {Marshall} \&
  {Fassnacht}}{{Nierenberg} et~al.}{2011}]{Nierenberg2011}
{Nierenberg} A.~M.,  {Auger} M.~W.,  {Treu} T.,  {Marshall} P.~J.,
  {Fassnacht} C.~D.,  2011, \apj, 731, 44

\bibitem[\protect\citeauthoryear{{Nierenberg}, {Auger}, {Treu}, {Marshall},
  {Fassnacht} \& {Busha}}{{Nierenberg} et~al.}{2012}]{Nierenberg2012}
{Nierenberg} A.~M.,  {Auger} M.~W.,  {Treu} T.,  {Marshall} P.~J.,  {Fassnacht}
  C.~D.,    {Busha} M.~T.,  2012, \apj, 752, 99

\bibitem[\protect\citeauthoryear{{Planck Collaboration}, {Ade}, {Aghanim},
  {Arnaud}, {Ashdown}, {Atrio-Barandela}, {Aumont}, {Baccigalupi}, {Balbi},
  {Banday} \& et al.}{{Planck Collaboration}
  et~al.}{2013}]{2013A&A...557A..52.}
{Planck Collaboration} {Ade} P.~A.~R.,  {Aghanim} N.,  {Arnaud} M.,  {Ashdown}
  M.,  {Atrio-Barandela} F.,  {Aumont} J.,  {Baccigalupi} C.,  {Balbi} A.,
  {Banday} A.~J.,    et al. 2013, AA, 557, A52

\bibitem[\protect\citeauthoryear{{Prescott} et~al.,}{{Prescott}
  et~al.}{2011}]{Prescott2011}
{Prescott} M.,  et~al., 2011, \mnras, 417, 1374

\bibitem[\protect\citeauthoryear{{Presotto}, {Girardi}, {Nonino}, {Mercurio},
  {Grillo}, {Rosati}, {Biviano} \& {Annunziatella}}{{Presotto}
  et~al.}{2014}]{Presotto2014}
{Presotto} V.,  {Girardi} M.,  {Nonino} M.,  {Mercurio} A.,  {Grillo} C.,
  {Rosati} P.,  {Biviano} A.,    {Annunziatella} M.,  2014, ArXiv e-prints

\bibitem[\protect\citeauthoryear{{Rocha}, {Peter} \& {Bullock}}{{Rocha}
  et~al.}{2012}]{Rocha2012}
{Rocha} M.,  {Peter} A.~H.~G.,    {Bullock} J.,  2012, \mnras, 425, 231

\bibitem[\protect\citeauthoryear{{Sales} \& {Lambas}}{{Sales} \&
  {Lambas}}{2005}]{Sales2005}
{Sales} L.,  {Lambas} D.~G.,  2005, \mnras, 356, 1045

\bibitem[\protect\citeauthoryear{{Sales}, {Navarro}, {Lambas}, {White} \&
  {Croton}}{{Sales} et~al.}{2007}]{Sales2007c}
{Sales} L.~V.,  {Navarro} J.~F.,  {Lambas} D.~G.,  {White} S.~D.~M.,
  {Croton} D.~J.,  2007, \mnras, 382, 1901

\bibitem[\protect\citeauthoryear{{Springel}, {Wang}, {Vogelsberger}, {Ludlow},
  {Jenkins}, {Helmi}, {Navarro}, {Frenk} \& {White}}{{Springel}
  et~al.}{2008}]{Springel2008}
{Springel} V.,  {Wang} J.,  {Vogelsberger} M.,  {Ludlow} A.,  {Jenkins} A.,
  {Helmi} A.,  {Navarro} J.~F.,  {Frenk} C.~S.,    {White} S.~D.~M.,  2008,
  \mnras, 391, 1685

\bibitem[\protect\citeauthoryear{{Springel}, {White}, {Jenkins}, {Frenk},
  {Yoshida}, {Gao}, {Navarro}, {Thacker}, {Croton}, {Helly}, {Peacock}, {Cole},
  {Thomas}, {Couchman}, {Evrard}, {Colberg} \& {Pearce}}{{Springel}
  et~al.}{2005}]{Springel2005a}
{Springel} V.,  {White} S.~D.~M.,  {Jenkins} A.,  {Frenk} C.~S.,  {Yoshida} N.,
   {Gao} L.,  {Navarro} J.,  {Thacker} R.,  {Croton} D.,  {Helly} J.,
  {Peacock} J.~A.,  {Cole} S.,  {Thomas} P.,  {Couchman} H.,  {Evrard} A.,
  {Colberg} J.,    {Pearce} F.,  2005, \nat, 435, 629

\bibitem[\protect\citeauthoryear{{Springel}, {White}, {Tormen} \&
  {Kauffmann}}{{Springel} et~al.}{2001}]{Springel2001}
{Springel} V.,  {White} S.~D.~M.,  {Tormen} G.,    {Kauffmann} G.,  2001,
  \mnras, 328, 726

\bibitem[\protect\citeauthoryear{{Tal}, {Wake} \& {van Dokkum}}{{Tal}
  et~al.}{2012}]{Tal2012}
{Tal} T.,  {Wake} D.~A.,    {van Dokkum} P.~G.,  2012, \apjl, 751, L5

\bibitem[\protect\citeauthoryear{{Trinh}, {Barton}, {Bullock}, {Cooper},
  {Zentner} \& {Wechsler}}{{Trinh} et~al.}{2013}]{Trinh2013}
{Trinh} C.~Q.,  {Barton} E.~J.,  {Bullock} J.~S.,  {Cooper} M.~C.,  {Zentner}
  A.~R.,    {Wechsler} R.~H.,  2013, \mnras, 436, 635

\bibitem[\protect\citeauthoryear{{van den Bosch}, {Aquino}, {Yang}, {Mo},
  {Pasquali}, {McIntosh}, {Weinmann} \& {Kang}}{{van den Bosch}
  et~al.}{2008}]{vandenBosch2008}
{van den Bosch} F.~C.,  {Aquino} D.,  {Yang} X.,  {Mo} H.~J.,  {Pasquali} A.,
  {McIntosh} D.~H.,  {Weinmann} S.~M.,    {Kang} X.,  2008, \mnras, 387, 79

\bibitem[\protect\citeauthoryear{{van den Bosch}, {Yang}, {Mo} \&
  {Norberg}}{{van den Bosch} et~al.}{2005}]{vandenBosch2005}
{van den Bosch} F.~C.,  {Yang} X.,  {Mo} H.~J.,    {Norberg} P.,  2005, \mnras,
  356, 1233

\bibitem[\protect\citeauthoryear{{Wang}, {De Lucia}, {Kitzbichler} \&
  {White}}{{Wang} et~al.}{2008}]{Wang2008}
{Wang} J.,  {De Lucia} G.,  {Kitzbichler} M.~G.,    {White} S.~D.~M.,  2008,
  \mnras, 384, 1301

\bibitem[\protect\citeauthoryear{{Wang}, {Frenk}, {Navarro}, {Gao} \&
  {Sawala}}{{Wang} et~al.}{2012}]{Wang2012}
{Wang} J.,  {Frenk} C.~S.,  {Navarro} J.~F.,  {Gao} L.,    {Sawala} T.,  2012,
  \mnras, p.~3369

\bibitem[\protect\citeauthoryear{{Wang}, {Li}, {Kauffmann} \& {De
  Lucia}}{{Wang} et~al.}{2007}]{WangLi2007}
{Wang} L.,  {Li} C.,  {Kauffmann} G.,    {De Lucia} G.,  2007, \mnras, 377,
  1419

\bibitem[\protect\citeauthoryear{{Wang}, {Jing}, {Li}, {Okumura} \&
  {Han}}{{Wang} et~al.}{2011}]{Wang2011}
{Wang} W.,  {Jing} Y.~P.,  {Li} C.,  {Okumura} T.,    {Han} J.,  2011, \apj,
  734, 88

\bibitem[\protect\citeauthoryear{{Wang} \& {White}}{{Wang} \&
  {White}}{2012}]{Wang_White2012}
{Wang} W.,  {White} S.~D.~M.,  2012, \mnras, 424, 2574

\bibitem[\protect\citeauthoryear{{Watson}, {Berlind}, {McBride}, {Hogg} \&
  {Jiang}}{{Watson} et~al.}{2012}]{Watson2012}
{Watson} D.~F.,  {Berlind} A.~A.,  {McBride} C.~K.,  {Hogg} D.~W.,    {Jiang}
  T.,  2012, \apj, 749, 83

\bibitem[\protect\citeauthoryear{{Weinmann}, {Kauffmann}, {van den Bosch},
  {Pasquali}, {McIntosh}, {Mo}, {Yang} \& {Guo}}{{Weinmann}
  et~al.}{2009}]{Weinmann2009}
{Weinmann} S.~M.,  {Kauffmann} G.,  {van den Bosch} F.~C.,  {Pasquali} A.,
  {McIntosh} D.~H.,  {Mo} H.,  {Yang} X.,    {Guo} Y.,  2009, \mnras, 394, 1213

\bibitem[\protect\citeauthoryear{{Weinmann}, {Kauffmann}, {von der Linden} \&
  {De Lucia}}{{Weinmann} et~al.}{2010}]{Weinmann2010}
{Weinmann} S.~M.,  {Kauffmann} G.,  {von der Linden} A.,    {De Lucia} G.,
  2010, \mnras, 406, 2249

\bibitem[\protect\citeauthoryear{{Weinmann}, {Lisker}, {Guo}, {Meyer} \&
  {Janz}}{{Weinmann} et~al.}{2011}]{Weinmann2011}
{Weinmann} S.~M.,  {Lisker} T.,  {Guo} Q.,  {Meyer} H.~T.,    {Janz} J.,  2011,
  \mnras, 416, 1197

\bibitem[\protect\citeauthoryear{{Weinmann}, {van den Bosch}, {Yang} \&
  {Mo}}{{Weinmann} et~al.}{2006}]{Weinmann2006a}
{Weinmann} S.~M.,  {van den Bosch} F.~C.,  {Yang} X.,    {Mo} H.~J.,  2006,
  \mnras, 366, 2

\bibitem[\protect\citeauthoryear{{Wetzel}, {Tinker} \& {Conroy}}{{Wetzel}
  et~al.}{2012}]{Wetzel2012}
{Wetzel} A.~R.,  {Tinker} J.~L.,    {Conroy} C.,  2012, \mnras, 424, 232

\bibitem[\protect\citeauthoryear{{Wetzel}, {Tinker}, {Conroy} \& {van den
  Bosch}}{{Wetzel} et~al.}{2013}]{Wetzel2013}
{Wetzel} A.~R.,  {Tinker} J.~L.,  {Conroy} C.,    {van den Bosch} F.~C.,  2013,
  \mnras, 432, 336

\bibitem[\protect\citeauthoryear{{Wheeler}, {Phillips}, {Cooper},
  {Boylan-Kolchin} \& {Bullock}}{{Wheeler} et~al.}{2014}]{Wheeler2014}
{Wheeler} C.,  {Phillips} J.~I.,  {Cooper} M.~C.,  {Boylan-Kolchin} M.,
  {Bullock} J.~S.,  2014, ArXiv e-prints

\bibitem[\protect\citeauthoryear{{White} \& {Rees}}{{White} \&
  {Rees}}{1978}]{White_Rees1978}
{White} S.~D.~M.,  {Rees} M.~J.,  1978, \mnras, 183, 341

\bibitem[\protect\citeauthoryear{{Wojtak} \& {Mamon}}{{Wojtak} \&
  {Mamon}}{2013}]{Wojtak2013}
{Wojtak} R.,  {Mamon} G.~A.,  2013, \mnras, 428, 2407

\bibitem[\protect\citeauthoryear{{Yang}, {Mo} \& {van den Bosch}}{{Yang}
  et~al.}{2009}]{Yang2009}
{Yang} X.,  {Mo} H.~J.,    {van den Bosch} F.~C.,  2009, \apj, 693, 830

\bibitem[\protect\citeauthoryear{{York} et~al.,}{{York}
  et~al.}{2000}]{York2000}
{York} D.~G.,  et~al., 2000, \aj, 120, 1579

\bibitem[\protect\citeauthoryear{{Zehavi}, {Zheng}, {Weinberg}, {Blanton},
  {Bahcall}, {Berlind}, {Brinkmann}, {Frieman}, {Gunn}, {Lupton}, {Nichol},
  {Percival}, {Schneider}, {Skibba}, {Strauss}, {Tegmark} \& {York}}{{Zehavi}
  et~al.}{2011}]{Zehavi2011}
{Zehavi} I.,  {Zheng} Z.,  {Weinberg} D.~H.,  {Blanton} M.~R.,  {Bahcall}
  N.~A.,  {Berlind} A.~A.,  {Brinkmann} J.,  {Frieman} J.~A.,  {Gunn} J.~E.,
  {Lupton} R.~H.,  {Nichol} R.~C.,  {Percival} W.~J.,  {Schneider} D.~P.,
  {Skibba} R.~A.,  {Strauss} M.~A.,  {Tegmark} M.,    {York} D.~G.,  2011,
  \apj, 736, 59

\bibitem[\protect\citeauthoryear{{Zhao}, {Jing}, {Mo} \& {B{\"o}rner}}{{Zhao}
  et~al.}{2009}]{Zhao2009}
{Zhao} D.~H.,  {Jing} Y.~P.,  {Mo} H.~J.,    {B{\"o}rner} G.,  2009, \apj, 707,
  354

\end{thebibliography}

\appendix
\section{Projection effects in a light-cone and the local environment of isolated galaxies}
\label{sec:appendix-a}

Throughout our paper we have compared rectilinear projections of
snapshots of the Millennium and Millennium-II simulations with
observations from SDSS.  Although most projection effects are taken
care of in a realistic way in these mock galaxy samples (redshifts are
computed using the line-of-sight distance within the box together with
peculiar velocities) there are several factors affecting the SDSS data
that are not properly represented. Among these, effects due to the
fixed flux limit of a real survey and the K-corrections needed to
obtain rest-frame magnitudes, fiber-fiber collision effects
which make it difficult to obtain redshifts for close galaxy pairs,
and the effects of the survey geometry stand out as possible sources
of concern for our analysis. To address these issues, we have
paralleled the analysis presented in the main body of our paper with
studies of satellite profiles obtained from a light-cone galaxy
catalogue taken from \citet{Henriques2012}
\footnote{This catalogue is available at
  http://www.mpa-garching.mpg.de/millennium.}.  This light-cone is
generated from the same Millennium-based galaxy population simulation
as our other model catalogues, but properly includes evolutionary and
band-pass shifting effects, as well as the flux limits of the survey,
a simplified model for the effects of fiber-fiber collision, and the
geometry of the SDSS mask. The selection for primaries and
satellites can be applied to this light-cone catalogue in {\it
  exactly} the same way as to the observed SDSS sample.

\begin{center} 
\begin{figure} 
\includegraphics[width=84mm]{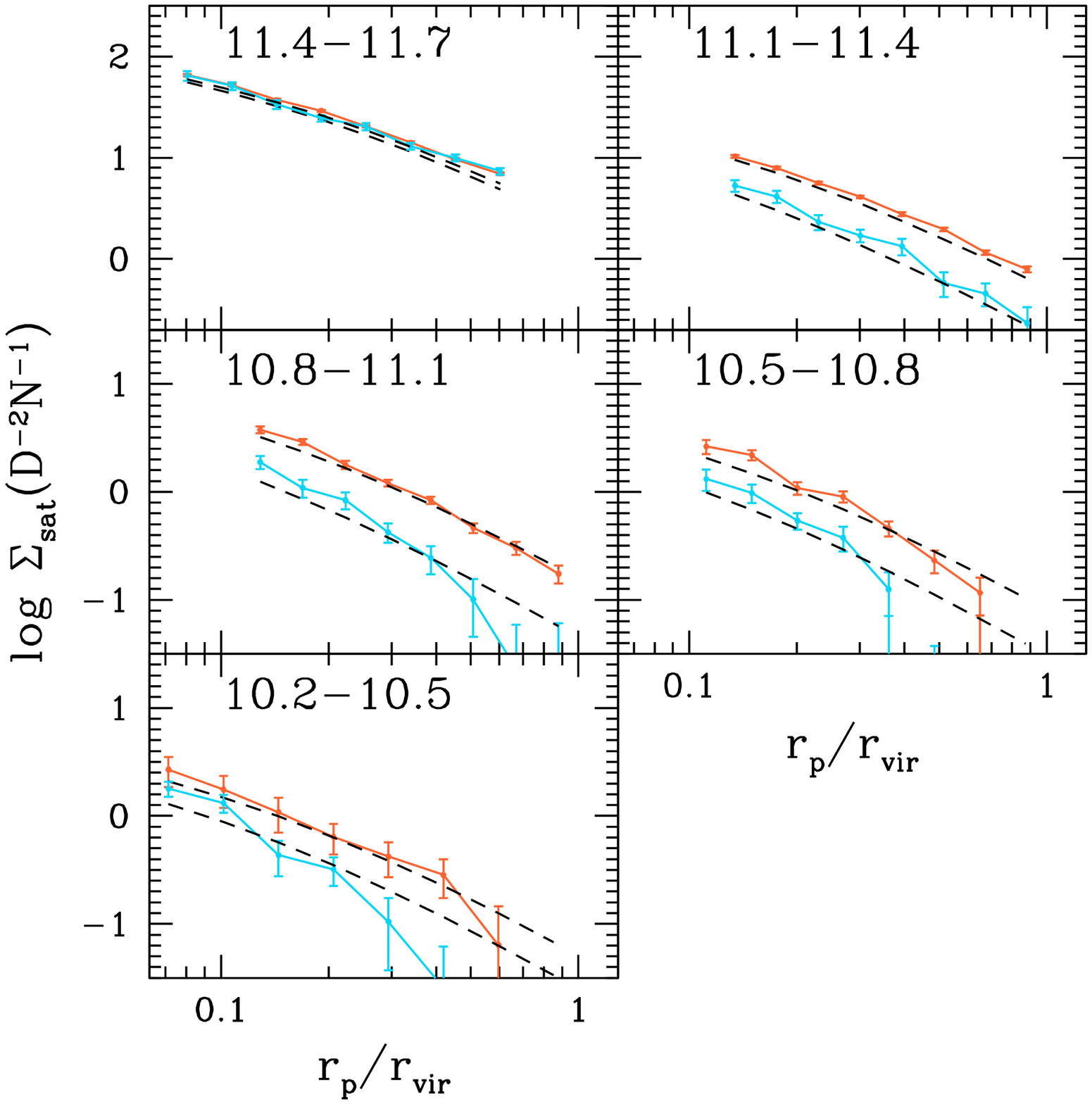} 
\caption{Same as Fig.~\ref{fig:densp_primcolor} but for a simulated
  light-cone that mimics observational effects in the SDSS (see
  Henriques et al. 2012). The black dashed 
  curves reproduce those in the right column of Fig.~\ref{fig:densp_primcolor}. 
  There is a good agreement between the results from the light-cone 
  and from the mock catalogue generated by rectilinear projection of 
  a simulation snapshot. The steep decline of satellite profiles at 
  large radii around blue low-mass primaries appears to reflect a bias 
  in the environment of our primaries (see the text for more details).}
\label{fig:densp_lightcone}
\end{figure}
\end{center}

Overall, we find good agreement between our original mock catalogue
and that generated from the light-cone. We illustrate this by showing
the projected number density profiles of satellites split according to
primary colour from the light-cone sample in
Fig.~\ref{fig:densp_lightcone}. Panels indicate different primary
stellar mass bins, with the $\log M_*/M_\odot$ ranges quoted at the
top left of each box. As before, orange and light-blue curves
correspond to satellites of red and blue primaries, respectively
(colour cuts for each primary stellar mass bin are as quoted in Table
1).  This figure should be compared with the right column of 
Fig.~\ref{fig:densp_primcolor}. To guide the eye the black 
dashed lines in Fig.~\ref{fig:densp_lightcone} are exactly those 
in the right column of Fig.~\ref{fig:densp_primcolor}. Agreement between 
the two samples is generally good, but for low-mass primaries,
particularly blue ones, satellite profiles tend to fall off more
rapidly at large radii than for the dark matter (black dashed lines).
This effect is only marginally detected but appears to reflect a bias
in the light-cone catalogue, since satellite profiles in 3D
(Fig.~\ref{fig:3Ddens}) and projected directly from the box (right
column Fig.~\ref{fig:densp_primcolor}) do not show such a steepening.

The probable source of this effect is an over-estimation of the
background subtraction in our light-cone samples (both observational
and simulated).  The isolation criteria for primaries used in this
paper and also in Paper I is quite strict, and as a result objects
that fulfill it are significantly biased towards low density
environments. Since our satellite profiles are estimated by
subtracting the {\it average} projected number counts from the full
catalogue, we over-correct for background at about the $\sim 1\%$
level.  Although this is not a large effect in the satellite
population as a whole, it becomes noticeable when the signal is weak,
i.e. at large distances from low-mass primaries.

\begin{center} 
\begin{figure} 
\includegraphics[width=84mm]{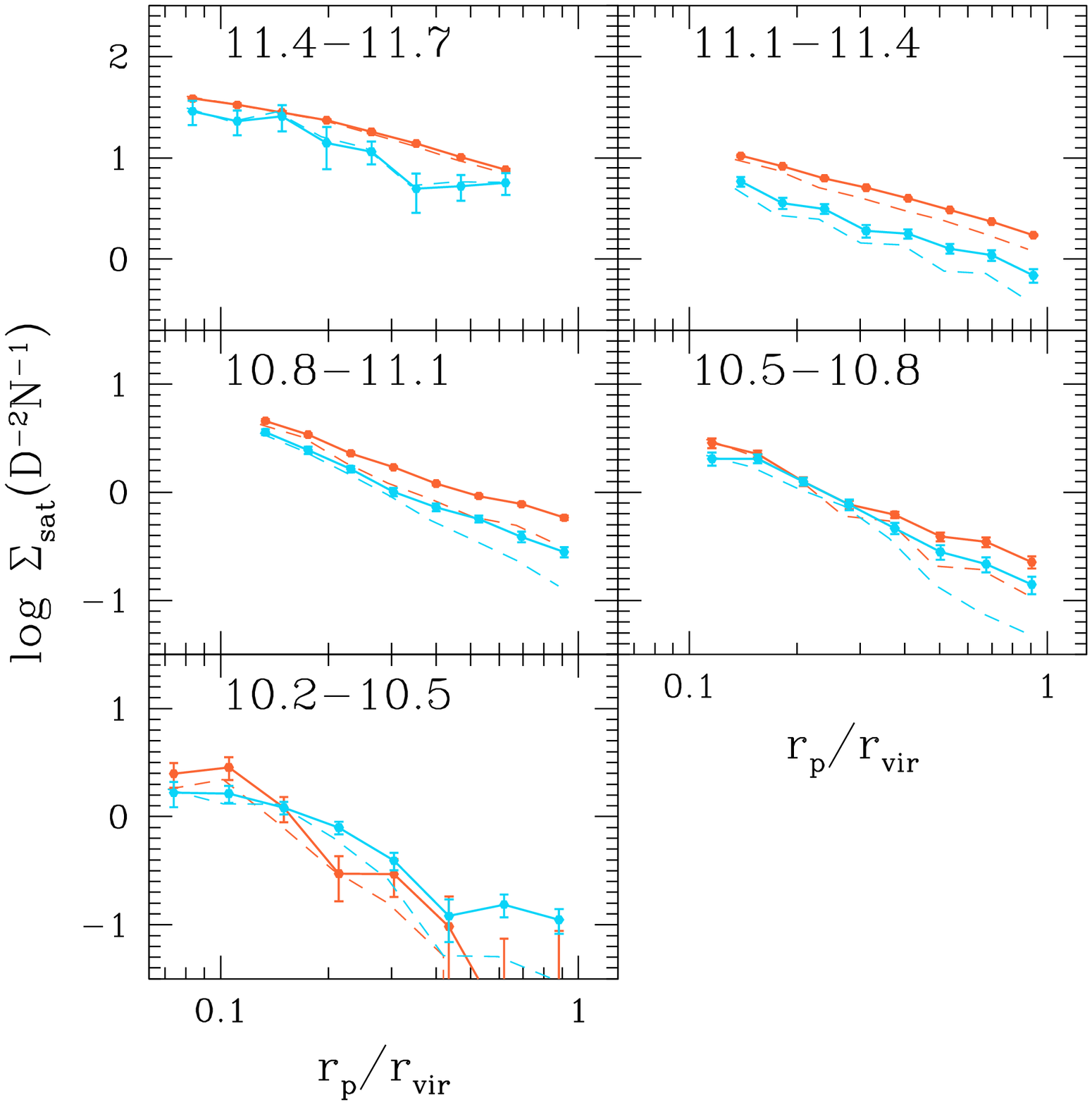} 
\caption{Satellite profiles for red and blue primary galaxies selected
  from SDSS with less strict isolation criteria (see the text for
  details). Thick orange and light-blue curves correspond to satellites 
  of red and blue primaries. The thin dashed orange and light-blue curves 
  reproduce those in the left column of Fig.~\ref{fig:densp_primcolor}. 
  The outer profiles of low-mass primaries become shallower with  
  less strict isolation.   }
\label{fig:densp_lbg}
\end{figure}
\end{center}

We explore the local environment of isolated galaxies further in 
Fig.\ref{fig:densp_lbg}, where we have
selected primary galaxies from SDSS with more relaxed isolation
criteria, increasing the primary sample by $\sim 40\%$ (primaries are
still required to be the brightest galaxy within 1~Mpc and the
line-of-sight velocity criterion is unchanged, but the stricter
isolation criterion within 500~kpc is eliminated, see
\citet{2013A&A...557A..52.}).  Thin dashed curves in
Fig.~\ref{fig:densp_lbg} reproduce those in the left column of
Fig.~\ref{fig:densp_primcolor}. The more relaxed isolation criteria
translate into significantly shallower satellite profiles at large
radii than those we obtained in Sec. 3.2 (the slope differences are
significant only for $r > 0.5 r_{\rm vir}$).  We have checked that
this excess is due to the inclusion of a larger fraction of primaries
that are not central galaxies of their FoF halo, with the consequences
for the satellite profiles discussed in Sec. 3.2. Interestingly, the
mean dark matter profiles in the simulation for this new primary sample 
show a similar excess at large radius, further evidence for our claim 
that satellites do indeed trace the underlying dark matter distribution 
very well, at least in the simulation. We also notice slightly higher 
normalizations for the inner satellite profiles in this new sample 
compared to Fig. 4. Primaries with relatively bright companions have 
more faint satellites than those which do not, an effect that we 
traced back to their having slightly more massive halos.

\end{document}